\documentclass{emulateapj}
\usepackage{amssymb,amsmath,amsthm}
\usepackage{longtable}
\tabletypesize{\scriptsize}
\usepackage{color}
\usepackage{pdfcolmk}
\definecolor{DarkGreen}{rgb}{0.0, 0.5, 0.0}
\definecolor{purple}{rgb}{0.6, 0.0, 0.4}

\newcommand{\degree}{^\circ}
\newcommand{\rms}{{\it rms}}
\newcommand{\snr}{{\it snr}}

\newcommand{\be}{\begin{equation}}
\newcommand{\ee}{\end{equation}}

\def\sigmaLim{5}
\def\noiseLim{0.75}
\def\nofs{{\it N(>\!S)}}
\def\specind#1#2{{#1/#2}}
\def\mason09{M09}

\begin{document}
\title{Radio Sources from a 31~GHz Sky Survey with the Sunyaev-Zel'dovich Array}

\author{
Stephen~Muchovej\altaffilmark{1,2},
Erik~Leitch\altaffilmark{3},
John~E.~Carlstrom\altaffilmark{3,4}, 
Thomas~Culverhouse\altaffilmark{3},
Chris~Greer\altaffilmark{3},
David~Hawkins\altaffilmark{1},
Ryan~Hennessy\altaffilmark{3},
Marshall~Joy\altaffilmark{5}, 
James~Lamb\altaffilmark{1}, 
Michael~Loh\altaffilmark{3},
Daniel~P.~Marrone\altaffilmark{3,6},
Amber~Miller\altaffilmark{7},
Tony~Mroczkowski\altaffilmark{2,8},
Clem~Pryke\altaffilmark{3},
Matthew~Sharp\altaffilmark{3},
David~Woody\altaffilmark{1}
}
\altaffiltext{1}{California Institute of Technology, Owens Valley Radio Observatory, Big Pine, CA 93513}
\altaffiltext{2}{Department of Astronomy, Columbia University, New York, NY 10027}
\altaffiltext{3}{Department of Astronomy and Astrophysics, Kavli Institute for Cosmological Physics, University of Chicago, Chicago, IL 60637}
\altaffiltext{4}{Dept. of Physics, Enrico Fermi Institute, University of Chicago, Chicago IL 60637}
\altaffiltext{5}{Space Sciences - VP62, NASA Marshall Space Flight Center, Huntsville, AL 35812}
\altaffiltext{6}{Jansky Fellow, National Radio Astronomy Observatory}
\altaffiltext{7}{Columbia Astrophysics Lab, Department of Physics, Columbia University, New York, NY}
\altaffiltext{8}{Dept. of Physics and Astronomy, U Penn, Philadelphia, PA}

\begin{abstract}

We present the first sample of 31-GHz selected sources to flux levels
of $1$~mJy.  From late 2005 to mid 2007, the Sunyaev-Zel'dovich Array
(SZA) observed 7.7 square degrees of the sky at 31~GHz to a median rms
of ${\rm 0.18 ~mJy/beam}$.  We identify 209 sources at greater than
5$\sigma$ significance in the 31~GHz maps, ranging in flux from
0.7~mJy to $\sim$ 200~mJy.  Archival NVSS data at 1.4~GHz and
observations at 5~GHz with the Very Large Array are used to
characterize the sources.  We determine the maximum-likelihood
integrated source count to be
${\it N(>\!S) = (27.2\pm 2.5)\deg^{-2} \times (S_{\it mJy})^{-1.18 \pm
    0.12}}$ over the flux range $0.7 - 15$~mJy.  This result is
significantly higher than predictions based on 1.4-GHz selected
samples, a discrepancy which can be explained by a small shift in the
spectral index distribution for faint 1.4~GHz sources.  From
comparison with previous measurements of sources within the central
arcminute of massive clusters, we derive an overdensity of ${\rm 6.8
  \pm 4.4}$, relative to field sources.

\end{abstract}
 
\keywords{techniques: interferometric, catalogs, surveys, cosmology:
cosmic microwave background, cosmology: observations, radio continuum:
general}

\section{Introduction}

Until recently, prohibitive integration times have limited deep
surveys at high radio frequency to small areas, or high flux
cutoffs. As a result, estimates of the properties of compact
extragalactic radio sources at frequencies above ${\rm \sim 15~GHz}$
have come primarily from observations at low-frequencies (in the $<$
10~GHz range), where the bulk of the sources are brighter,
extrapolated to higher frequencies by targeted followup campaigns
\citep{deZotti2005}.  While an understanding of high-frequency source
populations is interesting in its own right, it is also important for
the current generation of CMB and SZ experiments, many of which
operate at frequencies $\gtrsim 30$~GHz, where the source population
is poorly characterized.

In recent years the development of broad-band correlators has made
deep surveys of significant areas of sky possible with interferometric
arrays operating at 31~GHz.  In this paper, we report results of a 7.7
square degree, 31-GHz sky survey with the Sunyaev-Zel'dovich Array
(SZA).  Although the primary goal of the SZA was to measure CMB
anisotropy and to search for galaxy clusters via their
Sunyaev-Zel'dovich (SZ) effect, it has also yielded the first catalog
of $\sim$mJy sources selected at 31~GHz.  This paper focuses on the results
of the SZA survey observations as they pertain to the population of
high-frequency (${\rm \sim 31~GHz}$) selected compact sources.

Several experiments, such as WMAP \citep{bennett2003}, DASI
\citep{kovac2002}, CBI \citep{mason2003}, and the VSA
\citep{cleary2005}, have characterized sources at 31~GHz brighter than
5~mJy.  The present work extends this characterization to the fainter
31~GHz sources.  
As these faint sources are at or near the noise level of high-frequency CMB experiments, they represent a 
serious contaminant which must be carefully modeled and statistically accounted for.
The OVRO/BIMA SZ group \citep{coble2007}
have surveyed ${\rm \sim 29}$-GHz sources associated with massive clusters at the
mJy level; the SZA survey allows a comparison of this highly-selected
population with the bulk properties of field sources at 31~GHz.  More
recently, the CBI collaboration has followed up over 3000
1.4~GHz-selected sources at 31~GHz \citep{mason2009} and used these
observations to predict the source population at 31~GHz. Comparison of
this prediction with the SZA measurement at 31~GHz provides a direct
test of the assumptions that underlie this extrapolation.

The paper is organized as follows: in \S \ref{sec:bkg} we present a
description of the instrument and of the SZA observations.  In \S
\ref{sec:vla} we describe follow-up observations performed with the
Very Large Array (VLA), while \S \ref{sec:sourceExtract} details the
algorithm used to extract source fluxes from the SZA survey data.  The
characteristics of the 31~GHz-selected sample of sources are presented
in \S \ref{sec:popCharacteristics}, followed by a discussion in \S
\ref{sec:discussion}.  Conclusions are presented in \S
\ref{sec:conclusions}.

\section{Sunyaev-Zel'dovich Array Observations}
\label{sec:bkg}
\subsection{The Sunyaev-Zel'dovich Array}

The Sunyaev-Zel'dovich Array is an interferometer designed
specifically for detecting and imaging the SZ effect in galaxy
clusters, and is located at the Owens Valley Radio Observatory (OVRO).
The SZA is equipped with an 8-GHz wideband correlator and sensitive
26-36~GHz and 85-115~GHz receivers.  In this paper, we present results
only from 26-36~GHz (hereafter 31~GHz) SZA observations.

The SZA consists of eight 3.5-meter antennas.  For the observations
presented here, six were arranged in a close-packed configuration
(yielding high brightness sensitivity on angular scales typical of
clusters of galaxies), and two outlier antennas provided long
baselines for sensitivity to compact objects.  The SZA can therefore
be thought of as two complementary interferometers: one with a typical
resolution of a few arcminutes (short antenna separations, or {\it
baselines}), and one with resolution of about 23 arcseconds (long
baselines).  For a more detailed discussion of array layout and
corresponding resolution, see \cite{muchovej2007}.

In the limit where sky curvature is negligible over the instrument's
field of view, the response of an interferometer on a single baseline,
known as a {\it visibility}, can be approximated by:
\begin{eqnarray}
 V(u,v) &=& \int\!\!\!\int_{-\infty}^{+\infty}\!\!\!A(l,m)I(l,m) e^{-2\pi i \{ul + vm\}}{dl\,dm},
\label{eq:vis2}
\end{eqnarray}
where $u$ and $v$ are the baseline lengths projected onto
the sky, $l$ and $m$ are direction cosines measured with respect to
the $(u,v)$ axes, $A(l,m)$ is the normalized antenna beam pattern,
and $I(l,m)$ is the sky intensity distribution.
Eq.~\ref{eq:vis2} is a two-dimensional Fourier transform, the
inverse of which is the image of the source intensity multiplied by
the primary beam pattern, known as a {\it dirty map} $I_D$:
\begin{eqnarray}
\label{eq:dirty_image}\nonumber
I_D(l,m) &\equiv& A(l,m)I(l,m)\\
&=& \int\!\!\!\int_{-\infty}^{+\infty}\!\!\!V(u,v) e^{2\pi i \{ul + vm\}}{du\,dv}.
\end{eqnarray}
In practice, an interferometer measures discrete Fourier modes,
and structure in the dirty map is convolved with a function which 
reflects this incomplete Fourier-space sampling. This 
function, called the {\it synthesized beam}, is equivalent to the point-spread
function for the interferometer.

It is clear from Eq.~\ref{eq:dirty_image} that the field of view for
each dirty map is limited to the size of the primary beam, $A(l,m)$,
namely ${\rm \sim 11.0\arcmin}$ (FWHM) for the SZA at band center
($30.938$~GHz).  To image larger areas, we use linear mosaicking to
stitch together nearby pointings which cover the region of interest
\citep[e.g.,][]{pearson2003}.  Note that although large areas of sky
can be surveyed in this manner, the resulting mosaics contain no
information on size scales larger than that probed by the shortest
baseline.

The results presented in this paper are derived from two large SZA
projects: one to measure primary CMB anisotropy \citep{sharp2009}, and
the other to survey for galaxy clusters via their SZ effect.  Observations were conducted differently for these two projects, as detailed in \S\ref{sec:obs_strategy}.

\subsection{Field Selection} 
\begin{figure}[h!]
\begin{center}
\includegraphics[width=3.5in]{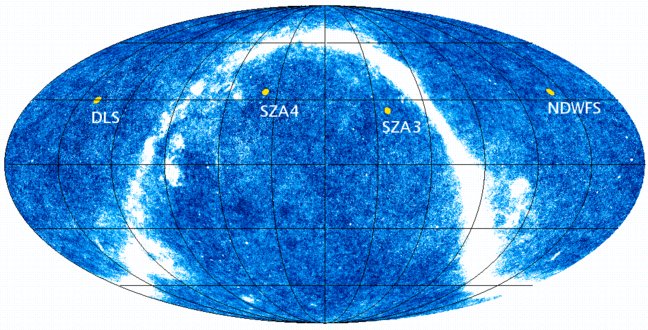}
\caption{IRAS ${\rm 100~\mu m}$ dust map with overlay of the SZA field locations.}
\label{fig:fields}
\end{center}
\end{figure}

The SZA survey fields were selected to lie far from the plane of the
Galaxy and to transit at high elevation at the OVRO site, minimizing
atmospheric noise while optimizing the imaging capabilities of
the array.  Fields were spaced equally in Right Ascension (RA) to
permit continuous observation. These constraints led to the selection
of four regions ranging in declination from ${\rm 25\degree}$ to ${\rm 35\degree}$.
Figure \ref{fig:fields} depicts the approximate locations of these
four fields overlaid on the IRAS ${\rm 100~\mu m}$ dust maps
\citep{iras}.

\subsection{Observation Strategy}
\label{sec:obs_strategy}

\subsubsection{Survey Mode} 

\begin{figure}[h!]
\begin{center}
\includegraphics[width=3.5in]{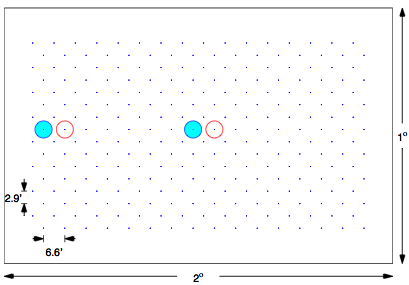}
\caption{Mosaic pointing locations for a given SZA survey field. The
fields are divided into 16 rows of 16 columns, with the pointings in
each row separated by 6.6\arcmin\ and each row offset from each other
by 2.9\arcmin.  This leads to each field being roughly 2 degrees by 1
degree in area.  In a single track the SZA observed four pointings
within a given row.  For example, pointings in the first and ninth,
followed by pointings in the second and tenth columns.}
\label{fig:trackStruct}
\end{center}
\end{figure}

Each of the four fields is split into 16 rows of 16 pointings.  The
pointings are equally spaced by 6.6\arcmin \ along great circles in
the RA direction, and each row is equally spaced by 2.9\arcmin \ in
the DEC direction.  Subsequent rows are offset from one another so
that that the first pointing in each row is shifted by 3.3\arcmin \ in
the RA direction relative to the previous row.  This means that for a
single field we observe an area that spans roughly 2 degrees in the RA
direction and 1 degree in the DEC direction (see Figure
\ref{fig:trackStruct}).

For each of the survey fields, data were taken daily in 6 hour {\it
tracks}.  In a single track, we observed two staggered pairs of
pointings, all within a single row.  These observations were performed
in a manner that permits ground subtraction from consecutive pointings
in a pair (although ground subtraction was not used in the analysis
presented here).  Each track results in roughly 1 hour of observation
on each of the four pointings, with very nearly the same Fourier
sampling for pairs of pointings.  A second track is run at a later
date, with the order of the pairs reversed, to ensure that the Fourier
sampling for all four pointings is comparable.  In Figure
\ref{fig:trackStruct} we show the position of the pointings in each
field, and indicate how the pointings were observed in a given track.

For each set of four pointings, this sequence is repeated three times over the span of roughly one year, so
that each pointing is observed in six total tracks, translating to
roughly 6 hours of observation per pointing over the duration of the
survey.

\subsubsection{CMB anisotropy Observations}

In addition to the survey observations, data were separately taken to
measure the anisotropy in the CMB.  These consisted of observations of
44 distinct pointings, each separated by one degree, which were not
mosaicked, but analyzed individually.  Of these 44 pointings, 11
overlap with pointings in the survey fields described above, and the
rest are within a two degree radius of the center of the four survey
fields.  Where they overlap, the analysis in this paper uses the
survey data only.  The track structure in the anisotropy observations
is similar to that in the survey analysis; see \cite{sharp2009} for
further details.

\subsection{Observations}

\begin{deluxetable*}{lcrrccccccc}[!h]
\tabletypesize{\scriptsize}
\tablecolumns{8}
\setlength{\tabcolsep}{1mm}
\tablecaption{Survey Observations}
\tablehead{
\colhead{Field Name}& \multicolumn{2}{c}
{\underline{Field Center (J2000)}}& \multicolumn{2}{c}{\underline{Calibrators}} & \colhead{Dates} & \colhead{Integration} & \colhead{Rows} \\
\colhead{} & \colhead{$\alpha$}& \colhead{$\delta$} & \colhead{Bandpass (Jy)\tablenotemark{a}} & \colhead{Gain (Jy)\tablenotemark{a}} & \colhead{of Observations} & \colhead{Time (hrs)} &\colhead{Covered}\\}
\startdata
SZA4   & 02$^h$15$^m$38$^s$.3 &32$^{\circ}$08$^{\prime}$21$^{\prime\prime}$ & J2253+161 (11.6) & J0237+288 (2.9) & 07/11/2006 to 07/25/2007 & 687 & 7\\
DLS    & 09$^h$19$^m$40$^s$.0 &30$^{\circ}$01$^{\prime}$26$^{\prime\prime}$ & J0319+415 (11.0) & J0854+201 (5.4) & 11/18/2005 to 07/06/2007 & 1054 & 14\\ 
NDWFS  & 14$^h$30$^m$08$^s$.0 &35$^{\circ}$08$^{\prime}$34$^{\prime\prime}$ & J1229+020 (25.3) & J1331+305 (2.1) & 11/19/2005 to 07/23/2007 & 1000 & 14\\ 
SZA3   & 21$^h$30$^m$07$^s$.0 &25$^{\circ}$01$^{\prime}$26$^{\prime\prime}$ & J1642+398 (\phantom{2}5.5) & J2139+143 (1.4) & 11/13/2005 to 07/25/2007 & 1245 & 16\\
\enddata
\label{tab:obsTable}
\tablenotetext{a}{Fluxes obtained from 31~GHz SZA observations of sources on April 16, 2006. }
\end{deluxetable*}

Table \ref{tab:obsTable} presents details of the mosaicked SZA survey
observations (see \cite{sharp2009} for the equivalent information on
the CMB anisotropy observations).  The second and third columns show
the approximate pointing center of each 16-row field.  We also present
the bandpass and gain calibrators in the next two columns, with their
fluxes as measured by the SZA.  In the fifth column we give the time
range over which observations were taken, with the caveat that
observations were not performed every day during that time span.  The
penultimate column lists the total unflagged integration time for data
used in the analysis, and the final column gives the number of rows
observed in each field.  To ensure uniform coverage of all fields,
tracks were repeated when necessary.  Note that the full 16 rows were
not observed for all fields, due to maintenance operations,
instrumental characterization, and RFI monitoring.
For the first 8 months of observations, the SZA4 field was used for
the dedicated CMB anisotropy measurements described above.  As a
result, only 7 rows in the SZA4 field were completed in survey mode.

The data in the SZA survey correspond to 1493 tracks taken between
November 13, 2005 and July 25, 2007.  The data in the CMB anisotropy
measurements correspond to an additional 414 tracks taken between
November 12, 2005 and October 25, 2007.  The analysis in this paper
refers to the full 1907 tracks taken in both observing modes.

Data for an individual track were calibrated using a suite of
MATLAB\footnote{The Mathworks, Version 7.0.4 (R14),
\tt{http://www.mathworks.com/products/matlab}} routines, which
constitute a complete pipeline for flagging, calibrating, and reducing
visibility data \citep{muchovej2007}.  Although the data were reduced
exactly as described in that paper, data collection differed in a few
key ways: four distinct pointings were observed before observing a
calibrator, and system temperature measurements were performed every
eight minutes.  The absolute flux calibration is referenced to Mars,
assuming the Rudy (1987) temperature model, and is estimated to be
accurate to better than 10\%.  Typical system temperatures measured
throughout the survey were in the range 40-50~K.  Flagging of the data
as described in \cite{muchovej2007} resulted in a loss of roughly 15\%
of the data.  At the end of a single 6-hour track, we achieved a noise
level of approximately 1.5~mJy/beam in each pointing of the short and
long baseline maps.

\subsection{Resulting Mosaics}
\label{sec:mosaics}

Once data on all pointings in a given field are reduced, we construct a linear mosaic of the field
on a regular grid of
3.3\arcsec\ resolution.  This scale is much less than the requirement
for Nyquist sampling of the data, $\frac{1}{2D_{max}}$, where
$D_{max}$ is the longest baseline, and ensures that the number of
pixels over 2 degrees is a power of 2, convenient for fast
inversion of the Fourier data via FFT.  The maps are a
result of combining the data across our 8~GHz of bandwidth, so that they
approximate the sky at the central observing frequency,
30.938~GHz.  The primary beam is calculated from the Fourier
transform of the aperture illumination of each telescope at the
central observing frequency, modelled as a Gaussian with a central
obscuration corresponding to the secondary mirror.  

Unlike the maps of individual pointings, the mosaicked maps are an
estimate of the true sky signal at each point in the map; that is, the
taper of the primary beam has been divided out.  
Due to the overlap of neighboring pointings, the effective noise is
approximately uniform in the interior of the mosaics, but increases
significantly towards the edge of the mosaicked images.  We limit the
survey area by applying an edge cutoff in our mosaicked maps where the
effective noise is $> \noiseLim$~mJy/beam (corresponding roughly to
the one-third power point of the beam, given the noise in a single
pointing).

In Table \ref{tab:sensTable} we show the noise properties of the
observed fields.  We present the minimum and median noise (in
mJy/beam) for mosaic maps made with long baselines only, short
baselines only, and with the combination of the two.  The median noise
is calculated only in the region within which the noise is less than
the \noiseLim~mJy/beam cutoff.  The last column indicates the total
area covered in each field.  That the minimum and median pixel noise
values are similar is an indication of the uniformity of the coverage
in the survey fields.  This is not the case in the CMB anisotropy
fields as they consisted of discrete pointings which do not overlap on
the sky.

\section{VLA Observations}
\label{sec:vla}

As described above, the primary goals of the SZA project were a
small-scale CMB anisotropy measurement and an SZ survey for clusters
of galaxies, both of which require an accurate accounting of
foreground emission.  Although the long and short baseline data
provide some intrinsic ability to discriminate compact objects, as
discussed in \S\ref{sec:bkg}, high-sensitivity follow-up observations
of the SZA fields were obtained with the VLA\footnote{The Very Large
Array is a facility of the National Radio Astronomy Observatory,
operated by Associated Universities, Inc., under a cooperative
agreement with the National Science Foundation.} to facilitate source
removal.

VLA data at 1.4~GHz are publicly available from the NRAO VLA Sky
Survey (NVSS; \citet{nvssSurvey}) for all of the SZA fields, but are
limited by the relatively coarse resolution of the NVSS, 45\arcsec\ FWHM, and high
detection threshold of 2.5~mJy.  The finer resolution (5\arcsec) and deeper
sensitivity (\rms\ of 0.15~mJy) obtained with the FIRST survey
\citep{firstSurvey} are better suited for this analysis, but data are only
available on half of the fields (namely the DLS and NDWFS fields).

To complement the NVSS and FIRST observations, we obtained
high-sensitivity VLA observations at 5~GHz.  Approximately 116 hours
of observation were required to cover all four of the SZA fields, in
four time blocks between 24 Feb 2007 and 15 April 2007.  Data were
taken at a center frequency of 4.86~GHz with the VLA D-array
configuration, by mosaicking 180 pointings in each field.  These pointings were
arranged in 9 rows of 20 pointings in a hexagonal pattern, equally
spaced by 6 arcminutes.

The pointings were mosaicked, after CLEANing, using the AIPS package
FLATN, with the noise calculated using RMSD.  Sources were extracted
using the AIPS SAD algorithm, which iteratively removes the brightest
point in a mosaic using an elliptical Gaussian model.  The {\it
rms} achieved on the VLA 5~GHz observations was roughly ${\rm
70\,\mu Jy/beam}$, resulting in 859 sources from all four fields,
down to a ${\rm 5\sigma}$ depth of ${\rm \sim 0.33~mJy}$.  For the
fields observed in CMB anisotropy mode, VLA follow-up at 8~GHz was
obtained.  These observations are described in \cite{sharp2009}.  In
this analysis, we use the VLA data to improve extraction of compact
sources from our data as described in \S \ref{sec:sourceExtract}.

\begin{deluxetable*}{lccccccc}[!h]
\tablewidth{0pt}
\tabletypesize{\scriptsize}
\tablecolumns{8}
\setlength{\tabcolsep}{2.3mm}
\tablecaption{Survey Sensitivity}
\tablehead{
\colhead{Field}& \multicolumn{2}{c}{\underline{Short Baselines}} & \multicolumn{2}{c}{\underline{Long Baselines}} & \multicolumn{2}{c}{\underline{All Baselines}} & \colhead{Area} \\
\colhead{Name} & \colhead{Minimum {\it rms}} & \colhead{Median {\it rms}} &  \colhead{Minimum {\it rms}} & \colhead{Median {\it rms}} &  \colhead{Minimum {\it rms}} & \colhead{Median {\it rms}} &  \colhead{Covered}\\
\colhead{} & \colhead{(mJy/beam)} & \colhead{(mJy/beam)} & \colhead{(mJy/beam)} & \colhead{(mJy/beam)} & \colhead{(mJy/beam)} & \colhead{(mJy/beam)} & \colhead{ (${\rm degree^2}$)}}
\startdata
SZA4       & 0.218 & 0.264 & 0.231 & 0.283 & 0.159 & 0.193 & 0.91 \\
DLS        & 0.202 & 0.238 & 0.200 & 0.251 & 0.142 & 0.173 & 1.57 \\
NDWFS      & 0.219 & 0.239 & 0.219 & 0.249 & 0.156 & 0.172 & 1.56 \\
SZA3       & 0.213 & 0.232 & 0.218 & 0.241 & 0.153 & 0.167 & 1.74 \\
CMB fields & 0.152 & 0.393 & 0.153 & 0.418 & 0.108 & 0.286 & 1.91 \\
\enddata
\label{tab:sensTable}
\end{deluxetable*}

\section{Source Extraction from the SZA survey}
\label{sec:sourceExtract}

\subsection{Overview}
\label{sec:iterativeAlgo}

Source
identification begins in the image plane, with inspection of the
combined (short and long baseline) significance (\snr) maps for the
brightest pixel with significance greater than \sigmaLim.
Once we identify the location of a source, we next determine whether
the source is extended or unresolved, as seen by the SZA or the VLA, and whether
this candidate is a single source, or a collection of nearby sources.
Due to the
complex sidelobe structure of the synthesized beam (see
\S\ref{sec:bkg}), nearby sources must be removed simultaneously
from the interferometric data; we therefore fit any additional
sources within 45\arcsec\ of the primary source location, roughly twice the
synthesized beam width of the long baseline maps.  

Once we have determined all sources near the brightest in the map
which are to be removed from the data, as well as their morphology
(compact/extended), we solve for source properties by fitting to the
multi-pointing visibility data.
For computational expediency, we
describe the sources as functions with analytic Fourier transforms
(see \S\ref{sec:models}).
The best-fit models are removed from the Fourier data, and the mosaics
are regenerated.  This process is repeated iteratively until there are
no sources brighter than ${\rm\sigmaLim\sigma}$ in the significance
maps.

\subsection{Models}
\label{sec:models}

When fitting sources in the mosaicked maps, we model the intensity
distribution of an unresolved source as a delta function,
\begin{equation}
\label{eq:Ips}
I(\nu,\vec{x})=I_o\left(\frac{\nu}{{\nu_o}}\right)^{-\alpha} \delta(\vec{x\
}-\vec{x_o}),
\end{equation}
where $\nu$ is the observation frequency, $I_o$ is the intensity at frequency ${\nu_o}$,
defined for the SZA as the
center frequency of observations, and $\alpha$ is the spectral index.

We model any extended source as a 2-dimensional Gaussian, i.e., up to a normalization,
\begin{equation}
\label{eq:extI}
I(\nu)\propto I_o\left(\frac{\nu}{{\nu_o}}\right)^{-\alpha}  e^{-\frac{ (l^\prime -{l_c}^\prime)^2}{2{\sigma_l}^2}}  e^{-\frac{ (m^\prime -{m_c}^\prime)^2}{2{\sigma_m}^2}},
\end{equation}
where $l_c$ and $m_c$ are the coordinates of the centroid, $\theta$ is the orientation angle, and $\sigma_l$ and
$\sigma_m$ are the FWHM of the semi-major and semi-minor axes, respectively.
In Eq.~\ref{eq:extI}, the primed coordinate system ($l^\prime$,
$m^\prime$) is related to the unprimed coordinate ($l, m$) via
a rotation by the orientation angle $\theta$.  

This model has an analytic Fourier transform provided we can take the
primary beam of the SZA to be constant over the extent of the source,
i.e., that the source is small compared to the primary beam.  In
practice, extended models are fit only to pointings within 6\arcmin\ of
the source, well within the FWHM of the primary beam.  For a source
1\arcmin\ in extent, a factor of $\sim3$ larger than the most extended
source in our data set, this assumption leads to an error in
the fitted flux that is below 5\%.

\subsection{Unresolved vs.\ Extended Sources}
\label{sec:psVsExt}
Sources in the SZA maps are cross-checked against the VLA 5~GHz
catalog, and a source is determined to have a counterpart at 5~GHz if
a source exists within 8\arcsec\ of the SZA location, a small fraction
of the highest resolution element of the SZA.  If a counterpart is
found, we consider the source extended if its size at 5~GHz is greater
than 22.5\arcsec (the SZA long-baseline synthesized beam FWHM).  In
the rare event that the source has no 5~GHz counterpart (due
presumably to source variability or resolution effects when observing
extended sources) we check within 8\arcsec\ of the SZA location in the
1.4~GHz NVSS or FIRST catalogs.  When no counterpart is found in any
of the catalogs (${\rm \sim 3\%\ }$ of sources), we compare the flux
at that location in the short and long baseline SZA maps to determine
if it is extended.  If the fluxes are consistent with each other the
source is classified as unresolved.

\subsection{Source Location Determination}
\label{sec:sourceLoc}
To minimize the number of degrees of freedom in each source fit, we
fix the location of SZA sources with VLA 
counterparts to the VLA positions.  
We have verified that the locations of the VLA counterparts agree with locations
fitted to the SZA data to within 3\arcsec, and indicate no
systematic pointing offsets.

Locations are not fixed for sources determined to be extended, and for
very bright sources at 31~GHz.  The dynamic range (ratio of the fitted
source flux to the post-fit residuals) in the mosaicked maps is
approximately 35, limited by the long-term pointing accuracy of the
instrument and uncertainties in the primary beam.
For sources brighter than 15~mJy, however, we can fit the locations
accurately in the individual pointings, resulting in an overall
dynamic range in the mosaics of 70, or a factor of 2 improvement.  We have
verified that agreement in the fitted source location between
pointings is at the arcsecond level.

\subsection{Spectral Index Fitting}

For source brighter than $\sim 3.5 $~mJy, the SZA's large bandwidth
allows for the simultaneous determination of the spectral index across
8~GHz.
For fainter sources, however, we cannot meaningfully constrain the spectral index of the
source from the SZA data alone.  To reduce the number of degrees of
freedom for these sources, we use the VLA 5~GHz observations (where available) as a
second frequency, fixing the spectral index to
\begin{equation}
\label{eq:alpha}
\alpha = -{\ln{\left(I_o/I_{\rm 5~GHz}\right)}}/{\ln{\left({\nu_o}/{\rm 5~GHz}\right)}}
\end{equation}
in Eq.~\ref{eq:Ips}.  Note that these spectral indices are used only to
reduce the residuals in the maps when extracting sources. Analysis of
the spectral index distribution of these sources by comparison with
low-frequency data is presented in \S\ref{sec:psalphas}.

\section{31~GHz Population Characteristics}
\label{sec:popCharacteristics}
\subsection{Source Sample}

Following the procedure outlined above, we identified 209 sources at
31~GHz with \snr\ greater than \sigmaLim.  From this sample, 40 of the
sources were determined to be extended in the VLA data, yet none were
seen to be resolved by the SZA, i.e. the FWHM of the fit Gaussians are
all smaller than 22.5\arcsec.  In Table~\ref{tab:srcs} of the Appendix we present the full list of
sources. The brightest source we detect has a flux of
${\rm 204}$~mJy.  Counterparts were found in the VLA 5~GHz
catalog for 162 of the 171 sources with 5~GHz coverage, and
1.4~GHz NVSS counterparts were found for 157 of the 209 sources.  All 1.4~GHz
sources in the regions for which we have 5~GHz coverage were also
found in the 5~GHz data.  VLA 8~GHz coverage is available in the
region for which we do not have 5~GHz coverage, and all 38 sources
detected at 31~GHz in those regions have 8~GHz counterparts
\citep{sharp2009}.  Of the 162 sources with 5~GHz counterparts, 34 were found
to have higher fluxes at 31~GHz than at 5~GHz.  In addition, 8 of the
157 sources with 1.4~GHz counterparts were found to be inverted.

\subsection{Source Count}
\label{sec:psdnds}

The differential source count per unit area as a function of flux,
$dN/dS$, is typically described by a power law, i.e.,
\begin{equation}
\label{eq:dnds_law}
\frac{dN}{dS} = N_0 \left(\frac{S}{S_0}\right)^{-\gamma},
\end{equation}
where $N_0$ is a normalization parameter per unit area, $S$ is the
source flux, and $\gamma$ is the power law index.  In the analysis, we
take $S_0 = 1$~mJy and express the normalization as a number of sources per 
square degree.  
As models of radio source populations suggest a break in the count
near 15~mJy, we fit a power-law only to sources below this limit,
excluding 7 sources from the previous list. Subsequent analysis of the
source count is performed on 202 sources.

\subsubsection{Power law index, $\gamma$}
\label{sec:gamma}

Given a set of $N_S$ observed source fluxes $\{S_i\}$, we can solve for the
underlying population parameters by maximizing the likelihood of the
dataset,
\begin{equation}
\mathcal{L} = \prod_i^{N_S} p(S_i|N_0, \gamma, \sigma_i),
\label{eq:prob}
\end{equation}
where $S_i$ is the observed flux of the source, and $\sigma_i$ is the
map noise at the source location.  In the presence of noise, an
observed source flux $S$ is the combination of a ``true'' flux $S_t$
and noise $N = S-S_t$.  The probability in Eq.~\ref{eq:prob} can
therefore be written as an integral over all possible pairs of $S_t$
and $N$ that will produce an observed flux $S_i$:
\begin{eqnarray}\nonumber
p(S_i|N_0,\gamma, \sigma_i) &\propto& \int_0^{\infty} p_S(S_t|N_0,\gamma)\,p_N(S_i-S_t|\sigma_i)\,dS_t \\
&\propto& \int_0^{\infty} N_0 {S_t}^{-\gamma} e^{-(S_i-S_t)^2/2\sigma_i^2}\,dS_t,
\label{eq:Pgammasig}
\end{eqnarray}
up to a constant of proportionality \citep[e.g.,][]{murdoch1973}. It
is clear that when $p(S_i|N_0,\gamma,\sigma_i)$ is 
normalized, $N_0$ drops out of the equation.  Thus, this method can
only be used to calculate the power law index $\gamma$; the {\it relative} frequency of sources of
different flux is related only to the shape parameter $\gamma$, and is
independent of the {\it total} number of sources.  In the noiseless
case, it can be trivially shown that the resulting estimator for
$\gamma$ meets the Fischer-Neyman criterion for a sufficient
statistic, i.e., that it utilizes all the information about the shape
parameter contained in the data set \citep[e.g.,][]{crawford1970}.

In the case of non-uniform noise, the above expression leads to a
normalized likelihood that depends in a complicated way on the
relative areas at different noise levels in the survey.  However,
because our sources are selected by \snr\ and not flux, it is
convenient to transform directly into the \snr\ basis, leading to an
expression for the likelihood that is independent of the noise, and
therefore of the relative areas at different noise levels:
\begin{equation}
p(s_i|\gamma) \propto \int_0^{\infty} {s_t}^{-\gamma} e^{-(s_i-s_t)^2/2}\,ds_t,
\label{eq:probsnr}
\end{equation}
where $s \equiv S/\sigma$.  We note in passing that while the kernel
of Eq.~\ref{eq:probsnr} formally diverges as $s_t \rightarrow 0$, this
is merely an artifact of the implicit assumption that the
single-source hypothesis dominates the probability, i.e., that we are
considering only sources well above the confusion limit.  This
assumption is consistent with the results quoted in
\S\ref{sec:results}, which indicate that the weakest source flux
included in the analysis is approximately 14 times the confusion
limit.  In practice, we truncate Eq.~\ref{eq:probsnr} at an $s_t$
which corresponds to a flux well above the confusion limit at the
lowest map noise, and below which the kernel contributes negligibly to
the integral.

We estimate the uncertainty in the calculated value of $\gamma$ by
using a quadratic approximation of the likelihood about its maximum.
We have verified through extensive simulation that this method
produces an unbiased estimate of the power law parameter $\gamma$ and
its associated uncertainty $\sigma_\gamma$.

\subsubsection{Normalization, $N_0$}

We can invoke Bayes' theorem to write the likelihood of a set of
parameters $\{N_0, \gamma\}$ given the data $D$ as
\begin{eqnarray}\nonumber
p(N_0,\gamma | D) &\propto& p(D|N_0, \gamma)\,p(N_0, \gamma) \\
&\propto& p(D|N_0, \gamma)\,p(N_0)\,p(\gamma).
\label{eq:normEq}
\end{eqnarray}
Assuming a uniform prior for $N_0$, $p(N_0) = {\it const}$, and integrating over $\gamma$, we have
\begin{equation}
p(N_0|D) \propto \int{p(D|N_0,\gamma)\,p(\gamma)\,d\gamma}.
\end{equation}
The distribution $p(\gamma)$ is just the likelihood derived in the
last section, or for computational convenience, its Gaussian
approximation.  The probability of the data given $N_0$ and $\gamma$
is the Poisson probability of the observed number of sources $N_S$,
given the expected number of sources $\lambda(N_0,\gamma)$,

\begin{eqnarray}\nonumber
p(D|N_0,\gamma) \propto {e^{-\lambda}\lambda^{N_S}\over{{N_S}!}},
\end{eqnarray}
where $\lambda(N_0,\gamma)$ is derived by integrating $dN/dS$ over the
noise mask of the SZA survey (given in Table~\ref{tab:area} of the Appendix).  The maximum likelihood estimate of the
normalization is then just the value of $N_0$ that maximizes
$p(N_0|D)$.

\subsubsection{Results}
\label{sec:results}

Placing a significance cutoff on the source detections at 5$\sigma$,
the source count follows a power law with $\gamma = 2.18 \pm 0.12$,
and a normalization at 1~mJy of $N_0 = 32.1 \pm 3.0$ deg$^{-2}$ (in
the $0.7 - 15$~mJy range).  Integrating Equation \ref{eq:dnds_law}
with these parameters yields an integrated source count of
\begin{equation}
\label{eq:intdnds}
\nofs = (27.2\pm 2.5)\deg^{-2} \times (S_{\it mJy})^{-1.18 \pm 0.12} 
\end{equation}

As discussed in \S\ref{sec:mosaics}, the effective noise in the
interior of the mosaicked maps is nearly uniform, but increases
rapidly towards the edges.  Because of this nonuniformity, the
$5\sigma$ significance threshold used to select sources independent of
noise level does not translate simply into a uniform completeness
limit across the full area of the survey. From the noise distribution, the sample is
expected to be $98\%$ complete above 1.4~mJy over a 4.3 square degree area, but
$98\%$ complete above 5~mJy over the full 7.7 square degree area.
Here, we define the 98\% completeness limit as the flux above which
the probability of missing a source due to noise fluctuations is $<
2\%$.

\subsection{Spectral Index Distribution}
\label{sec:psalphas}

We calculate spectral indices for all sources in the sample (including
those brighter than 15~mJy) with counterparts in the 5~GHz or 1.4~GHz
VLA catalogs, where the spectral index $\alpha$ is defined as in
Eq.~\ref{eq:Ips}.  
For each source, we construct the probability distribution for the
spectral index by sampling from the flux distributions at each
frequency.  In the presence of noise, the conditional probability of
the ``true'' 31~GHz flux $S_t$, given an observed flux $S$ is
\begin{equation}
p(S_t|S,\gamma,\sigma) \propto p_S(S_t|\gamma)\,p_N(S-S_t|\sigma),
\label{eq:boost}
\end{equation}
which is just the kernel of Eq.~\ref{eq:Pgammasig}.  For $\gamma > 0$,
$p_S(S_t|\gamma)$ increases as $S_t \rightarrow 0$, and the observed
flux $S$ is therefore generally ``boosted'' by the presence of noise.
Marginalizing this kernel over $\gamma$ using the likelihood derived
in \S\ref{sec:gamma} yields the distribution of $S_t$.  For the VLA
fluxes, no correction for boosting has been made.  This process is
repeated for each source, and the resulting probabilities are co-added
to form the spectral index distribution for all sources with
counterparts.

\begin{figure}
\begin{center}
\includegraphics[width=3.5in]{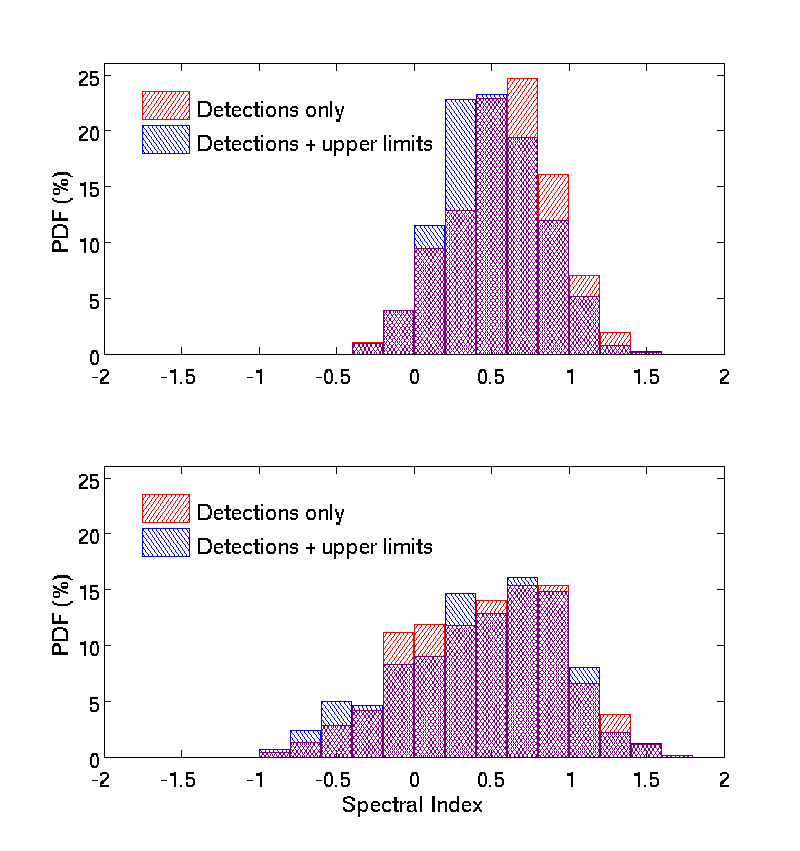}
\caption[Normalized spectral index distribution for 31~GHz-selected
sources detected with the SZA]{{\it Top:} Spectral index distribution
for 31~GHz-selected sources detected with the SZA relative to their
1.4~GHz flux seen by NVSS.  Red histograms denote sources with
identified counterparts while blue histograms include upper limits for
undetected sources, assuming the NVSS completeness limit of 3.5~mJy.
{\it Bottom:} Same histograms but for counterparts found in the 5~GHz
VLA follow-up data, with a limiting flux of 0.33~mJy at 5~GHz.}
\label{fig:alpha_fig}
\end{center}
\end{figure}

By correlating with the NVSS catalog, as described in
\S\ref{sec:psVsExt}, we find 7 instances where the higher resolution
of the SZA and VLA 5~GHz data, compared to NVSS, results in the
detection of two sources associated with a single NVSS source.  When
calculating \specind{1.4}{31}~GHz spectral indices, we treat these
seven pairs as a single source.  As a result, we calculate
\specind{1.4}{31}~GHz for 150 out of 202 sources, shown in the top
panel of Figure~\ref{fig:alpha_fig}.  The distribution peaks at a
spectral index of ${\rm \alpha \sim 0.7}$, consistent with standard
synchrotron emission, as expected for most sources bright at 1.4~GHz.
We note that of the sources with 1.4~GHz counterparts, approximately
5\% are inverted, i.e., have ${\rm \alpha < 0}$. Accounting for
sources undetected at 1.4~GHz will likely increase this fraction. For
these sources, we use the NVSS completeness limit (${\rm \sim
3.5~mJy}$) to construct upper limits for their spectral indices.
Figure~\ref{fig:alpha_fig} illustrates the shift toward more inverted
spectral indices when these limits are included.

We repeat the same calculation with the 162 sources with 5~GHz
counterparts, noting that only 171 of the 209 sources are covered by
the 5~GHz follow-up.  The resulting spectral index distribution for
these 162 sources is shown in the bottom panel of Figure
\ref{fig:alpha_fig}.  The distribution peaks at a value of ${\rm
\alpha \sim 0.8}$.  Of these 162 sources, we find that 14\% are
inverted.  Incorporating limits (${S_{\rm 5~GHz} \sim 0.33}$~mJy) for
nine sources which were not detected at 5~GHz has a minimal effect on
the distribution, shifting it marginally towards more inverted
sources.

We stress that information about the intrinsic spectral index
distribution of this population of sources can only be inferred from
the observed distributions by properly accounting for the selection
thresholds at each frequency.  For example, we can find particular
choices of intrinsic spectral index distribution that reproduce both
distributions shown in Figure~\ref{fig:alpha_fig} (though in
general we have no reason to expect a single distribution to describe
both the \specind{1.4}{31}\ and \specind{5}{31}~GHz spectral indices).
We therefore refrain from drawing any conclusions about frequency
evolution in spectral indices from these data.

\section{Discussion}

\label{sec:discussion}

In Figure \ref{fig:dnds_all}, we plot the SZA $dN/dS$
measurement over the range $0.7-15$~mJy.  In the top panel of the
figure, we show results from 31~GHz-selected
sources reported by OVRO/BIMA \citep{coble2007}, CBI
\citep{mason2003}, DASI \citep{kovac2002}, and the VSA
\citep{cleary2005}.  In the bottom panel we plot the projection from
\cite{deZotti2005}, and from \cite{mason2009}, based on targeted
31-GHz followup of a 1.4~GHz-selected sample of sources from the NVSS
catalog.

\subsection{Comparison with Cluster Source Measurements}

The only prior measurement of 31-GHz selected sources complete to a
comparable flux level is the \cite{coble2007} sample of $\sim 100$
sources brighter than 1~mJy, found serendipitously in targeted
observations of massive clusters of galaxies.  Since the vast majority
of these sources lie within the inner regions of clusters, these
measurements permit a direct comparison of the cluster population with
the 31-GHz field-source population (this work).  By integrating the
field-source count from the SZA and the cluster source count from
\cite{coble2007} over the flux range $0.7-15$~mJy, we calculate the
overdensity of radio sources within the central arcminute of clusters
to be a factor ${\rm 6.8 \pm 4.4}$, consistent with their estimate.

\subsection{Comparison with Field Source Measurements}

\begin{figure}
\begin{center}
\includegraphics[width=3.5in]{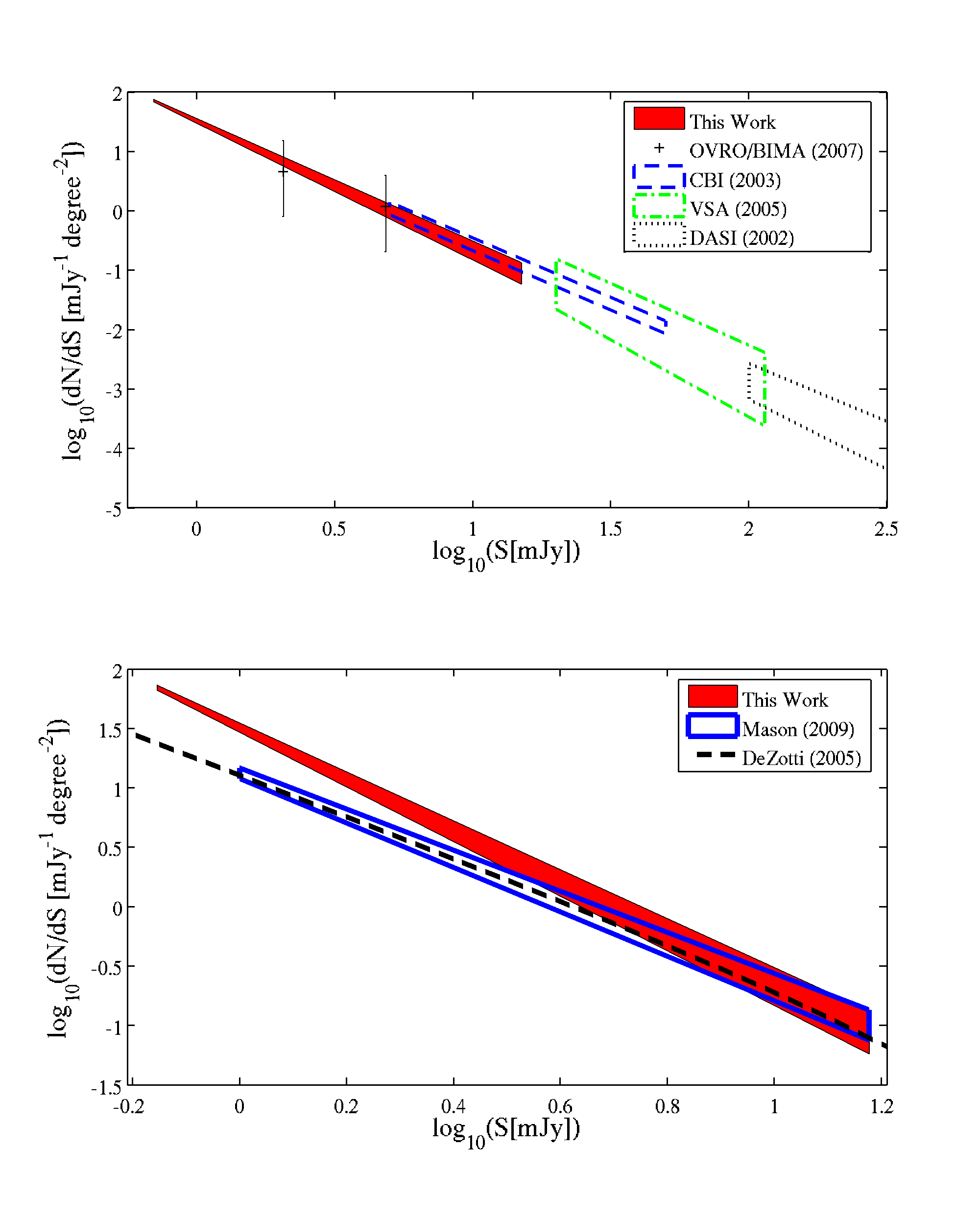}
\caption[Comparison of SZA derived dN/dS to theoretical predictions
and prior observations]{{\it Top:} Measurements of the 31~GHz dN/dS
from this work and prior experimental results from OVRO/BIMA
\citep{coble2007}, CBI \citep{mason2003}, the VSA \citep{cleary2005},
and DASI \citep{kovac2002}.  {\it Bottom:} Comparison of the SZA dN/dS
to projections from lower frequencies by \cite{deZotti2005} and
\cite{mason2009}.}
\label{fig:dnds_all}
\end{center}
\end{figure}

From the top panel of Figure \ref{fig:dnds_all}, we note the general
agreement between the SZA result and prior 31-GHz measurements, at the
higher completeness levels of those experiments (${\rm \gtrsim
4~mJy}$).  Agreement with the field-source count of \cite{coble2007}
is likewise good, although their sample in non-cluster fields consists
of only four sources.  From the bottom panel of Figure
\ref{fig:dnds_all}, we see that the SZA result is also consistent with
the projections \cite{deZotti2005} and \cite{mason2009} (hereafter
\mason09) for fluxes greater than ${\rm \sim 4~mJy}$.  At low flux,
however, they deviate significantly.

The \mason09\ projection is from a study of 3165 sources selected from
the NVSS catalog at 1.4~GHz and reobserved at the same central
observing frequency as the SZA.  Adopting the 1.4~GHz source
distribution of \cite{hopkins2003}, and an intrinsic
\specind{1.4}{31}~GHz spectral index distribution inferred from their
own observations, they estimate the integrated 31-GHz source count
over the same flux range as the SZA results to be ${\nofs = (16.7\pm
1.7) \deg^{-2} \times (S_{\it mJy})^{-0.80 \pm 0.07}}$.  This is
inconsistent with the count we determine directly at 31~GHz
(\S\ref{sec:results}) in both the normalization and the power law
index, as shown in the bottom panel of Figure~\ref{fig:dnds_all}.

As a consistency check, we generate simulated source populations
under the set of assumptions outlined in \mason09, apply the source
extraction algorithm detailed in \S\ref{sec:iterativeAlgo} to the
simulated data, and apply the formalism of \S\ref{sec:psdnds} to
calculate the expected $\nofs$ at 31~GHz.  In particular, we generate
a list of 1.4-GHz source fluxes over the range $25~{\rm \mu Jy} -
1~{\rm Jy}$ using the source count of \cite{hopkins2003},
and assigning spectral indices according to \mason09 to extrapolate
these fluxes to 31~GHz.  Sources are then assigned random locations on
a noise map identical to the actual SZA survey coverage, with
appropriate Gaussian noise added to the 31~GHz fluxes.  We select only
sources which would have been detected by the SZA, i.e., with \snr\
greater than 5, and calculate the integrated source count.  This
procedure is repeated for 100 realizations of source populations,
resulting in ${\nofs = (16.1^{+3.3}_{-2.9}) \deg^{-2} \times (S_{\it
mJy})^{-0.86 \pm0.18}}$.  This result is consistent with the
prediction of \mason09, and demonstrates that our methodology is
robust to the experimental details of the SZA survey, or differences
in source selection between the two measurements.

We have also investigated whether this discrepancy can be due to
differences in resolution between the SZA and the VLA in its NVSS
configuration.  However, of the 209 sources detected with the SZA, we
note that in only 7 cases are multiple sources close enough to appear as a
single source in the NVSS beam.  Treating these cases as single
sources, we obtain ${\nofs = (26.1\pm 2.3) \deg^{-2} \times (S_{\it
mJy})^{-1.19\pm 0.13}}$, an insignificant change from our nominal
result (Eq.~\ref{eq:intdnds}).

To reconcile the two measurements, we postulate a change in the
\specind{1.4}{31}~GHz spectal index distribution for sources below the
1.4~GHz flux limit of the \mason09\ study.  Although \mason09\ find no
significant difference between the spectral index distributions of
1.4~GHz sources with flux $> 10$~mJy and sources with flux $< 10$~mJy,
we note that their analysis is necessarily limited to sources brighter
than the NVSS completeness limit of 3.5~mJy.  Sources below this
limit, however, constitute a large fraction of the 1.4~GHz source
population used to predict the 31-GHz $dN/dS$, and small changes in
the assumed spectral index distribution of this population can have a
significant impact on the source population at 31~GHz.  We note also
that the SZA data suggest this hypothesis {\it ab initio}; while the
\mason09\ distribution predicts that fewer than $10\%$ of sources
detected by the SZA at 31~GHz would lack 1.4~GHz counterparts
$>3.5$~mJy, fully $25\%$ of SZA sources lack counterparts in the NVSS
catalog.

To test this hypothesis, we repeat the simulations described above,
using the \mason09\ spectral index distribution for sources above the
NVSS completeness limit (3.5~mJy), and a separate spectral index
distribution for sources with 1.4~GHz fluxes below 3.5~mJy.  For the dim 
sources, we assume a distribution whose shape is identical to that of
\mason09, but shifted with respect to it.  A good match to our data
can be obtained by shifting the \mason09\ spectral distribution by
0.35 to more inverted spectra, resulting in a predicted ${ \nofs =
(25.5\pm 2.9)\deg^{-2} \times (S_{\it mJy})^{-1.06 \pm 0.14}}$.  This
distribution has a peak spectral index of $\sim0.7$, and predicts that
${\rm \sim 7\%}$ of dim sources are inverted between 1.4 and 31~GHz.
This ad hoc model also accounts for the observed fraction of SZA
sources that fall below the NVSS detection threshold.  In addition,
this model provides a better fit than the single-component model to
the observed \specind{1.4}{31}~GHz spectral index distribution shown
in Figure~\ref{fig:alpha_fig}. Note that while this ansatz agrees
well with our data, a physically motivated model of faint source
spectra is likely to be much more complex.

Followup studies of faint 1.4-GHz sources at higher frequency have
indeed found evidence for a flattening of the spectral index distribution
near the mJy level.  
\cite{prandoni2006} find that for sources whose 1.4~GHz flux is
greater than 4~mJy, no sources are inverted between 1.4 and 5~GHz, but
that for sources dimmer than 4~mJy at 1.4~GHz, roughly 10\% of sources
are inverted, comparable to the fraction of inverted sources we find
between 1.4 and 31~GHz.  In addition, the study of \cite{donnelly1987}
conclude that the median \specind{1.4}{5}~GHz spectral index is
roughly 0.75 down to a 1.4~GHz flux of 0.25~mJy, and that the fraction
of flat spectrum sources increases from 22\% at $>0.5$~mJy to 41\% in
the $0.25-0.5$~mJy range.

\section{Conclusions}
\label{sec:conclusions}
We present a sample of 31~GHz-selected sources from a 7.7~square
degree survey obtained with the Sunyaev-Zel'dovich Array.
We identify 209 sources at $> 5\sigma$ detection significance, ranging
in flux from $0.7-204$~mJy.  A maximum likelihood determination of
the integrated source count results in 
${ \nofs = (27.2\pm 2.5)\deg^{-2} \times (S_{\it mJy})^{-1.18 \pm
    0.12}}$ in the flux range $0.7-15$~mJy.  Comparison with a
measurement of 31~GHz sources towards massive galaxy clusters leads to
an overdensity of ${\rm 6.8 \pm 4.4}$, for sources within the central
arcminute of massive clusters, relative to field sources.

Of the existing source samples selected at 31~GHz, the SZA sample
represents the only unbiased study of sources valid to $<1$~mJy.  This
sample will therefore be useful in refining source models that are
currently constrained by measurements at much higher flux.  Since
sources are also the most significant foreground for small-scale
temperature anisotropy experiments, these data will be useful for the
accurate inference of cosmological information from the current
generation of centimeter-wave CMB anisotropy and SZ cluster survey
measurements.

\acknowledgements We thank John Cartwright, Ben Reddall and Marcus
Runyan for their significant contributions to the construction and
commissioning of the SZA instrument.  We thank the staff of the Owens
Valley Radio Observatory and CARMA for their outstanding support.  We
thank Jonathan Sievers and Brian Mason for helpful discussions, Bryan
Butler, Claire Chandler, Gustaaf van Moorsel, and Meri Stanley for
their assistance with EVLA mosaicking observations, and Bryan Butler
and Mark Gurwell for providing the Mars model to which the SZA data
are calibrated.  We gratefully acknowledge the James S.\ McDonnell
Foundation, the National Science Foundation and the University of
Chicago for funding to construct the SZA.  The operation of the SZA is
supported by NSF Division of Astronomical Sciences through grant
AST-0604982. Partial support is provided by NSF Physics Frontier
Center grant PHY-0114422 to the Kavli Institute of Cosmological
Physics at the University of Chicago, and by NSF grants AST-0507545
and AST-05-07161 to Columbia University.  AM acknowledges support from
a Sloan Fellowship, and SM from an NSF
Astronomy and Astrophysics Fellowship, and CG, SM, and MS from NSF
Graduate Research Fellowships.


{\it Facilities:} \facility{SZA}, \facility{VLA}

\bibliographystyle{apj}

\bibliography{ms}

\begin{thebibliography}{18}
\expandafter\ifx\csname natexlab\endcsname\relax\def\natexlab#1{#1}\fi

\bibitem[{{Bennett} {et~al.}(2003){Bennett}, {Hill}, {Hinshaw}, {Nolta},
  {Odegard}, {Page}, {Spergel}, {Weiland}, {Wright}, {Halpern}, {Jarosik},
  {Kogut}, {Limon}, {Meyer}, {Tucker}, \& {Wollack}}]{bennett2003}
{Bennett}, C.~L., {Hill}, R.~S., {Hinshaw}, G., {Nolta}, M.~R., {Odegard}, N.,
  {Page}, L., {Spergel}, D.~N., {Weiland}, J.~L., {Wright}, E.~L., {Halpern},
  M., {Jarosik}, N., {Kogut}, A., {Limon}, M., {Meyer}, S.~S., {Tucker}, G.~S.,
  \& {Wollack}, E. 2003, \apjs, 148, 97

\bibitem[{{Cleary} {et~al.}(2005){Cleary}, {Taylor}, {Waldram}, {Battye},
  {Dickinson}, {Davies}, {Davis}, {Genova-Santos}, {Grainge}, {Jones},
  {Kneissl}, {Pooley}, {Rebolo}, {Rubi{\~n}o-Mart{\'{\i}}n}, {Saunders},
  {Scott}, {Slosar}, {Titterington}, \& {Watson}}]{cleary2005}
{Cleary}, K.~A., {Taylor}, A.~C., {Waldram}, E., {Battye}, R.~A., {Dickinson},
  C., {Davies}, R.~D., {Davis}, R.~J., {Genova-Santos}, R., {Grainge}, K.,
  {Jones}, M.~E., {Kneissl}, R., {Pooley}, G.~G., {Rebolo}, R.,
  {Rubi{\~n}o-Mart{\'{\i}}n}, J.~A., {Saunders}, R.~D.~E., {Scott}, P.~F.,
  {Slosar}, A., {Titterington}, D., \& {Watson}, R.~A. 2005, \mnras, 360, 340

\bibitem[{{Clegg}(1980)}]{iras}
{Clegg}, P.~E. 1980, \physscr, 21, 678

\bibitem[{{Coble} {et~al.}(2007){Coble}, {Bonamente}, {Carlstrom}, {Dawson},
  {Hasler}, {Holzapfel}, {Joy}, {La Roque}, {Marrone}, \& {Reese}}]{coble2007}
{Coble}, K., {Bonamente}, M., {Carlstrom}, J.~E., {Dawson}, K., {Hasler}, N.,
  {Holzapfel}, W., {Joy}, M., {La Roque}, S., {Marrone}, D.~P., \& {Reese},
  E.~D. 2007, \aj, 134, 897

\bibitem[{{Condon} {et~al.}(1998){Condon}, {Cotton}, {Greisen}, {Yin},
  {Perley}, {Taylor}, \& {Broderick}}]{nvssSurvey}
{Condon}, J.~J., {Cotton}, W.~D., {Greisen}, E.~W., {Yin}, Q.~F., {Perley},
  R.~A., {Taylor}, G.~B., \& {Broderick}, J.~J. 1998, \aj, 115, 1693

\bibitem[{{Crawford} {et~al.}(1970){Crawford}, {Jauncey}, \&
  {Murdoch}}]{crawford1970}
{Crawford}, D.~F., {Jauncey}, D.~L., \& {Murdoch}, H.~S. 1970, \apj, 162, 405

\bibitem[{{de Zotti} {et~al.}(2005){de Zotti}, {Ricci}, {Mesa}, {Silva},
  {Mazzotta}, {Toffolatti}, \& {Gonz{\'a}lez-Nuevo}}]{deZotti2005}
{de Zotti}, G., {Ricci}, R., {Mesa}, D., {Silva}, L., {Mazzotta}, P.,
  {Toffolatti}, L., \& {Gonz{\'a}lez-Nuevo}, J. 2005, \aap, 431, 893

\bibitem[{{Donnelly} {et~al.}(1987){Donnelly}, {Partridge}, \&
  {Windhorst}}]{donnelly1987}
{Donnelly}, R.~H., {Partridge}, R.~B., \& {Windhorst}, R.~A. 1987, \apj, 321,
  94

\bibitem[{{Hopkins} {et~al.}(2003){Hopkins}, {Afonso}, {Chan}, {Cram},
  {Georgakakis}, \& {Mobasher}}]{hopkins2003}
{Hopkins}, A.~M., {Afonso}, J., {Chan}, B., {Cram}, L.~E., {Georgakakis}, A.,
  \& {Mobasher}, B. 2003, \aj, 125, 465

\bibitem[{{Kovac} {et~al.}(2002){Kovac}, {Leitch}, {Pryke}, {Carlstrom},
  {Halverson}, \& {Holzapfel}}]{kovac2002}
{Kovac}, J.~M., {Leitch}, E.~M., {Pryke}, C., {Carlstrom}, J.~E., {Halverson},
  N.~W., \& {Holzapfel}, W.~L. 2002, \nat, 420, 772

\bibitem[{{Mason} {et~al.}(2003){Mason}, {Pearson}, {Readhead}, {Shepherd},
  {Sievers}, {Udomprasert}, {Cartwright}, {Farmer}, {Padin}, {Myers}, {Bond},
  {Contaldi}, {Pen}, {Prunet}, {Pogosyan}, {Carlstrom}, {Kovac}, {Leitch},
  {Pryke}, {Halverson}, {Holzapfel}, {Altamirano}, {Bronfman}, {Casassus},
  {May}, \& {Joy}}]{mason2003}
{Mason}, B.~S., {Pearson}, T.~J., {Readhead}, A.~C.~S., {Shepherd}, M.~C.,
  {Sievers}, J., {Udomprasert}, P.~S., {Cartwright}, J.~K., {Farmer}, A.~J.,
  {Padin}, S., {Myers}, S.~T., {Bond}, J.~R., {Contaldi}, C.~R., {Pen}, U.,
  {Prunet}, S., {Pogosyan}, D., {Carlstrom}, J.~E., {Kovac}, J., {Leitch},
  E.~M., {Pryke}, C., {Halverson}, N.~W., {Holzapfel}, W.~L., {Altamirano}, P.,
  {Bronfman}, L., {Casassus}, S., {May}, J., \& {Joy}, M. 2003, \apj, 591, 540

\bibitem[{{Mason} {et~al.}(2009){Mason}, {Weintraub}, {Sievers}, {Bond},
  {Myers}, {Pearson}, {Readhead}, \& {Shepherd}}]{mason2009}
{Mason}, B.~S., {Weintraub}, L., {Sievers}, J., {Bond}, J.~R., {Myers}, S.~T.,
  {Pearson}, T.~J., {Readhead}, A.~C.~S., \& {Shepherd}, M.~C. 2009, \apj, 704,
  1433

\bibitem[{{Muchovej} {et~al.}(2007){Muchovej}, {Mroczkowski}, {Carlstrom},
  {Cartwright}, {Greer}, {Hennessy}, {Loh}, {Pryke}, {Reddall}, {Runyan},
  {Sharp}, {Hawkins}, {Lamb}, {Woody}, {Joy}, {Leitch}, \&
  {Miller}}]{muchovej2007}
{Muchovej}, S., {Mroczkowski}, T., {Carlstrom}, J.~E., {Cartwright}, J.,
  {Greer}, C., {Hennessy}, R., {Loh}, M., {Pryke}, C., {Reddall}, B., {Runyan},
  M., {Sharp}, M., {Hawkins}, D., {Lamb}, J.~W., {Woody}, D., {Joy}, M.,
  {Leitch}, E.~M., \& {Miller}, A.~D. 2007, \apj, 663, 708

\bibitem[{{Murdoch} {et~al.}(1973){Murdoch}, {Crawford}, \&
  {Jauncey}}]{murdoch1973}
{Murdoch}, H.~S., {Crawford}, D.~F., \& {Jauncey}, D.~L. 1973, \apj, 183, 1

\bibitem[{{Pearson} {et~al.}(2003){Pearson}, {Mason}, {Readhead}, {Shepherd},
  {Sievers}, {Udomprasert}, {Cartwright}, {Farmer}, {Padin}, {Myers}, {Bond},
  {Contaldi}, {Pen}, {Prunet}, {Pogosyan}, {Carlstrom}, {Kovac}, {Leitch},
  {Pryke}, {Halverson}, {Holzapfel}, {Altamirano}, {Bronfman}, {Casassus},
  {May}, \& {Joy}}]{pearson2003}
{Pearson}, T.~J., {Mason}, B.~S., {Readhead}, A.~C.~S., {Shepherd}, M.~C.,
  {Sievers}, J.~L., {Udomprasert}, P.~S., {Cartwright}, J.~K., {Farmer}, A.~J.,
  {Padin}, S., {Myers}, S.~T., {Bond}, J.~R., {Contaldi}, C.~R., {Pen}, U.-L.,
  {Prunet}, S., {Pogosyan}, D., {Carlstrom}, J.~E., {Kovac}, J., {Leitch},
  E.~M., {Pryke}, C., {Halverson}, N.~W., {Holzapfel}, W.~L., {Altamirano}, P.,
  {Bronfman}, L., {Casassus}, S., {May}, J., \& {Joy}, M. 2003, \apj, 591, 556

\bibitem[{{Prandoni} {et~al.}(2006){Prandoni}, {Parma}, {Wieringa}, {de
  Ruiter}, {Gregorini}, {Mignano}, {Vettolani}, \& {Ekers}}]{prandoni2006}
{Prandoni}, I., {Parma}, P., {Wieringa}, M.~H., {de Ruiter}, H.~R.,
  {Gregorini}, L., {Mignano}, A., {Vettolani}, G., \& {Ekers}, R.~D. 2006,
  \aap, 457, 517

\bibitem[{{Sharp} {et~al.}(2009){Sharp}, {Marrone}, {Carlstrom}, {Culverhouse},
  {Greer}, {Hawkins}, {Hennessy}, {Joy}, {Lamb}, {Leitch}, {Loh}, {Miller},
  {Mroczkowski}, {Muchovej}, {Pryke}, \& {Woody}}]{sharp2009}
{Sharp}, M.~K., {Marrone}, D.~P., {Carlstrom}, J.~E., {Culverhouse}, T.,
  {Greer}, C., {Hawkins}, D., {Hennessy}, R., {Joy}, M., {Lamb}, J., {Leitch},
  E., {Loh}, M., {Miller}, A.~D., {Mroczkowski}, T., {Muchovej}, S., {Pryke},
  C., \& {Woody}, D. 2009, ArXiv e-prints

\bibitem[{{White} {et~al.}(1997){White}, {Becker}, {Helfand}, \&
  {Gregg}}]{firstSurvey}
{White}, R.~L., {Becker}, R.~H., {Helfand}, D.~J., \& {Gregg}, M.~D. 1997,
  \apj, 475, 479

\end{thebibliography}

\appendix
\begin{longtable}{ccrrrrr}
\tabletypesize{\scriptsize}
\tablecolumns{7}
\tablewidth{10pt}
\tablecaption{SZA 30~GHz Sources}
\tablehead{
\colhead{$\alpha$} & \colhead{$\delta$} & \colhead{${\rm {S_t}_{31~GHz}}$\tablenotemark{a}} & \colhead{$\rm{S_{31~GHz}}$\tablenotemark{b}} & \colhead{$\snr$\tablenotemark{c}} & \colhead{$\rm{S_{5~GHz}}$\tablenotemark{d}} & \colhead{$\rm{S_{1.4~GHz}}$\tablenotemark{e}}\\
 \colhead{(J2000)} & \colhead{(J2000)} & \colhead{(mJy)} & \colhead{(mJy)} & \colhead{} & \colhead{(mJy)} & \colhead{(mJy)}
 }
\tablenotetext{a}{Flux corrected for boosting (i.e., $S_t$ in Eq.~\ref{eq:boost}).  Errors do not reflect uncertainties in the absolute flux calibration}
\tablenotetext{b}{Best-fit flux, not accounting for boosting (i.e., $S$ in Eq.~\ref{eq:boost}).}
\tablenotetext{c}{Significance of detection in the composite mosaics}
\tablenotetext{d}{VLA 5~GHz flux.  Map \rms~ is 68$\mu$Jy/beam}
\tablenotetext{e}{1.4~GHz Flux from NVSS catalog}
\tablenotetext{$\dagger$}{Single source in NVSS catalog.  31 GHz fluxes were combined in the \specind{1.4}{31}~GHz spectral index analysis}
 \startdata
            &             &                          &      &       &      &       \\
09:16:37.50 & 29:50:35.53 & {$14.60^{+0.15}_{-0.15}$} & 14.60 & 93.80 & 31.0 & 51.1\\ 
09:18:01.17 & 30:01:28.00 & {$8.70^{+0.15}_{-0.17}$} & 8.70 & 55.13 & 32.0 & 93.7 \\ 
09:21:30.36 & 30:06:01.78 & {$6.43^{+0.17}_{-0.18}$} & 6.44 & 37.54 & 28.7 & 64.4 \\ 
09:20:28.42 & 29:47:57.00 & {$4.38^{+0.15}_{-0.16}$} & 4.39 & 30.12 & 19.0 & 10.2 \\ 
09:22:35.52 & 29:53:48.95 & {$4.08^{+0.14}_{-0.15}$} & 4.09 & 28.25 & 4.1 & {$<3.5$} \\ 
09:18:33.50 & 30:15:56.65 & {$3.64^{+0.17}_{-0.18}$} & 3.66 & 27.89 & 17.1 & 66.8\tablenotemark{$\dagger$} \\ 
09:18:32.51 & 30:16:09.89 & {$2.45^{+0.17}_{-0.18}$} & 2.48 & 24.46 & 9.2 & 66.8\tablenotemark{$\dagger$} \\ 
09:23:29.98 & 30:11:07.68 & {$4.58^{+0.19}_{-0.20}$} & 4.60 & 23.68 & 18.8 & 35.5 \\ 
09:22:01.24 & 30:14:11.86 & {$3.75^{+0.16}_{-0.17}$} & 3.77 & 23.64 & 3.1 & 34.5 \\ 
09:21:05.11 & 30:00:34.09 & {$3.44^{+0.15}_{-0.16}$} & 3.46 & 22.59 & 18.4 & 52.8 \\ 
09:23:31.36 & 29:49:27.80 & {$3.29^{+0.17}_{-0.18}$} & 3.31 & 17.85 & 5.4 & 48.4\tablenotemark{$\dagger$}\\ 
09:23:31.88 & 29:49:14.78 & {$1.54^{+0.17}_{-0.19}$} & 1.58 & 19.84 & 14.6 & 48.4\tablenotemark{$\dagger$} \\ 
09:20:37.26 & 29:56:26.93 & {$2.53^{+0.15}_{-0.16}$} & 2.55 & 16.59 & 5.8 & 14.8 \\ 
09:18:11.73 & 29:50:16.22 & {$3.12^{+0.15}_{-0.17}$} & 3.14 & 14.38 & 15.3 & 63.7 \\ 
09:16:17.99 & 29:40:40.40 & {$6.08^{+0.40}_{-0.41}$} & 6.14 & 14.07 & 12.3 & 24.8 \\ 
09:19:00.02 & 30:21:53.50 & {$2.41^{+0.20}_{-0.21}$} & 2.45 & 12.80 & 0.9 & {$<3.5$} \\ 
09:17:58.65 & 29:55:31.69 & {$1.74^{+0.14}_{-0.15}$} & 1.77 & 12.19 & 4.6 & 8.0 \\ 
09:19:11.46 & 30:07:33.91 & {$1.93^{+0.17}_{-0.19}$} & 1.97 & 11.43 & 4.2 & 8.2 \\ 
09:15:32.97 & 30:15:56.24 & {$2.77^{+0.26}_{-0.27}$} & 2.83 & 10.74 & 4.2 & {$<3.5$} \\ 
09:16:40.53 & 30:13:06.15 & {$1.85^{+0.16}_{-0.18}$} & 1.88 & 10.87 & 0.8 & {$<3.5$} \\ 
09:19:52.47 & 29:56:25.35 & {$1.07^{+0.15}_{-0.17}$} & 1.12 & 10.01 & 4.6 & 20.9 \\ 
09:19:40.43 & 29:59:32.82 & {$1.48^{+0.16}_{-0.16}$} & 1.52 & 9.99 & 7.6 & 25.2 \\ 
09:16:51.27 & 29:41:24.43 & {$3.32^{+0.34}_{-0.36}$} & 3.39 & 9.83 & 3.1 & {$<3.5$} \\ 
09:18:11.97 & 29:58:36.63 & {$1.36^{+0.15}_{-0.15}$} & 1.40 & 9.61 & 1.5 & 4.1 \\ 
09:16:25.35 & 29:52:16.26 & {$1.11^{+0.15}_{-0.16}$} & 1.15 & 8.68 & 3.2 & 11.0 \\ 
09:21:27.65 & 29:58:11.85 & {$1.46^{+0.14}_{-0.16}$} & 1.49 & 8.44 & 3.8 & 10.9 \\ 
09:23:39.94 & 30:06:30.24 & {$1.74^{+0.22}_{-0.23}$} & 1.80 & 8.54 & 12.6 & 39.7 \\ 
09:20:18.31 & 29:48:10.68 & {$1.10^{+0.15}_{-0.16}$} & 1.15 & 8.19 & 0.4 & {$<3.5$} \\ 
09:20:41.00 & 30:10:12.89 & {$1.27^{+0.16}_{-0.18}$} & 1.32 & 8.15 & 2.4 & 4.1 \\ 
09:18:57.85 & 30:21:07.24 & {$1.56^{+0.19}_{-0.21}$} & 1.61 & 7.67 & 5.6 & 32.1 \\ 
09:18:38.44 & 30:18:32.21 & {$1.29^{+0.17}_{-0.18}$} & 1.34 & 8.01 & 4.8 & 11.6 \\ 
09:19:29.27 & 29:59:57.34 & {$1.23^{+0.15}_{-0.17}$} & 1.28 & 7.96 & 8.2 & 8.7 \\ 
09:21:04.44 & 30:20:31.25 & {$1.32^{+0.17}_{-0.18}$} & 1.37 & 8.00 & 2.9 & 4.2 \\ 
09:16:26.72 & 29:59:23.77 & {$1.45^{+0.15}_{-0.16}$} & 1.49 & 7.76 & 1.2 & {$<3.5$} \\ 
09:17:01.04 & 30:01:26.81 & {$1.57^{+0.16}_{-0.17}$} & 1.61 & 7.63 & 4.9 & 16.5 \\ 
09:23:45.88 & 29:43:09.36 & {$2.48^{+0.34}_{-0.36}$} & 2.58 & 6.77 & 14.6 & 47.9 \\ 
09:23:34.35 & 30:12:05.13 & {$1.45^{+0.20}_{-0.21}$} & 1.51 & 7.09 & 1.5 & {$<3.5$} \\ 
09:22:06.74 & 30:28:58.82 & {$4.36^{+0.66}_{-0.68}$} & 4.57 & 6.79 & 6.7 & 3.5 \\ 
09:23:03.47 & 29:59:37.21 & {$0.98^{+0.15}_{-0.16}$} & 1.03 & 6.72 & {$<0.33$}& {$<3.5$} \\ 
09:20:50.45 & 30:17:35.77 & {$1.03^{+0.16}_{-0.18}$} & 1.09 & 6.38 & 4.6 & 10.3 \\ 
09:16:49.95 & 30:11:56.08 & {$0.93^{+0.17}_{-0.19}$} & 1.00 & 6.51 & {$<0.33$}& {$<3.5$} \\ 
09:15:49.44 & 29:50:02.20 & {$0.94^{+0.16}_{-0.17}$} & 1.00 & 6.11 & 3.1 & 5.9 \\ 
09:18:14.89 & 30:13:44.34 & {$0.97^{+0.16}_{-0.19}$} & 1.03 & 6.18 & 4.9 & 14.1 \\ 
09:23:19.44 & 29:46:07.68 & {$0.97^{+0.18}_{-0.20}$} & 1.05 & 5.58 & 0.6 & {$<3.5$} \\ 
09:22:16.60 & 30:15:07.99 & {$1.04^{+0.16}_{-0.17}$} & 1.10 & 5.52 & 1.0 & 3.6 \\ 
09:22:01.25 & 29:56:01.58 & {$0.76^{+0.14}_{-0.16}$} & 0.81 & 5.42 & 1.0 & {$<3.5$} \\ 
09:17:06.56 & 29:50:23.21 & {$0.76^{+0.16}_{-0.17}$} & 0.84 & 5.43 & 4.5 & 16.8 \\ 
21:34:12.54 & 25:20:05.00 & {$100.74^{+0.39}_{-0.41}$} & 100.74 & 243.60 & 145.0 & 123.4 \\ 
21:28:19.99 & 25:06:39.95 & {$13.68^{+0.17}_{-0.18}$} & 13.68 & 60.65 & 55.1 & 126.5 \\ 
21:27:19.43 & 25:12:50.59 & {$4.42^{+0.17}_{-0.17}$} & 4.44 & 39.38 & 4.3 & 68.8\tablenotemark{$\dagger$}\\ 
21:27:18.69 & 25:13:12.58 & {$3.04^{+0.16}_{-0.18}$} & 3.05 & 34.28 & 16.2 & 68.8\tablenotemark{$\dagger$}\\ 
21:27:18.44 & 25:12:20.60 & {$1.67^{+0.17}_{-0.17}$} & 1.71 & 24.01 & 9.1 & 2.1 \\ 
21:33:04.62 & 25:20:47.99 & {$6.83^{+0.18}_{-0.19}$} & 6.85 & 37.60 & 15.7 & 33.2 \\ 
21:28:52.38 & 24:39:06.47 & {$8.16^{+0.23}_{-0.24}$} & 8.17 & 35.33 & 9.8 & 4.8 \\ 
21:28:53.25 & 25:20:21.47 & {$3.41^{+0.17}_{-0.19}$} & 3.43 & 19.31 & 6.9 & 11.1 \\ 
21:32:08.45 & 25:12:09.99 & {$2.86^{+0.15}_{-0.17}$} & 2.88 & 18.25 & 3.6 & 4.5 \\ 
21:26:39.91 & 25:03:49.19 & {$2.37^{+0.16}_{-0.17}$} & 2.39 & 14.72 & 4.2 & {$<3.5$} \\ 
21:33:57.20 & 25:15:56.97 & {$2.82^{+0.21}_{-0.21}$} & 2.85 & 13.96 & 2.6 & 4.4 \\ 
21:32:30.18 & 25:02:11.37 & {$2.07^{+0.16}_{-0.17}$} & 2.09 & 12.57 & 5.5 & 15.1 \\ 
21:31:53.81 & 24:39:33.05 & {$2.39^{+0.20}_{-0.21}$} & 2.43 & 11.81 & 6.7 & 10.1 \\ 
21:30:44.02 & 24:49:33.86 & {$2.05^{+0.15}_{-0.17}$} & 2.08 & 10.59 & 7.8 & 20.8 \\ 
21:29:57.88 & 25:07:25.01 & {$1.53^{+0.16}_{-0.17}$} & 1.57 & 9.31 & 4.0 & 7.4 \\ 
21:27:21.49 & 24:58:17.25 & {$1.55^{+0.15}_{-0.17}$} & 1.59 & 9.47 & 1.1 & {$<3.5$} \\ 
21:27:00.95 & 24:52:26.77 & {$1.49^{+0.15}_{-0.17}$} & 1.53 & 9.35 & 1.6 & {$<3.5$} \\ 
21:33:55.85 & 25:12:29.07 & {$1.92^{+0.20}_{-0.21}$} & 1.97 & 9.42 & 3.3 & 2.5 \\ 
21:33:36.18 & 25:06:17.37 & {$1.38^{+0.16}_{-0.17}$} & 1.43 & 9.08 & 8.6 & 30.6 \\ 
21:27:55.18 & 25:01:44.83 & {$1.38^{+0.16}_{-0.17}$} & 1.42 & 8.75 & 1.4 & {$<3.5$} \\ 
21:33:10.62 & 25:15:56.43 & {$1.34^{+0.16}_{-0.17}$} & 1.38 & 8.12 & 2.5 & 5.0 \\ 
21:33:01.83 & 24:53:25.04 & {$1.30^{+0.16}_{-0.17}$} & 1.34 & 8.09 & 4.9 & 3.1 \\ 
21:31:47.69 & 25:18:27.96 & {$1.31^{+0.16}_{-0.17}$} & 1.35 & 8.05 & 4.0 & 9.1 \\ 
21:27:12.18 & 25:10:37.86 & {$1.16^{+0.16}_{-0.17}$} & 1.22 & 7.50 & 10.9 & 39.9 \\ 
21:32:52.46 & 24:55:34.07 & {$1.53^{+0.16}_{-0.17}$} & 1.56 & 6.72 & 2.7 & 8.7 \\ 
21:28:09.99 & 24:41:36.19 & {$1.28^{+0.18}_{-0.19}$} & 1.34 & 7.06 & 1.8 & {$<3.5$} \\ 
21:31:17.47 & 24:42:33.05 & {$1.00^{+0.17}_{-0.19}$} & 1.06 & 6.78 & {$<0.33$}& {$<3.5$} \\ 
21:31:29.21 & 24:38:37.85 & {$1.34^{+0.23}_{-0.24}$} & 1.43 & 6.64 & 4.3 & 11.0 \\ 
21:27:36.49 & 24:38:47.79 & {$1.49^{+0.23}_{-0.23}$} & 1.56 & 6.53 & 1.9 & 3.0 \\ 
21:28:51.49 & 24:58:10.41 & {$0.97^{+0.16}_{-0.18}$} & 1.03 & 5.89 & 1.7 & 3.3 \\ 
21:28:44.81 & 25:06:25.11 & {$0.99^{+0.16}_{-0.18}$} & 1.05 & 6.26 & 1.7 & {$<3.5$} \\ 
21:28:13.46 & 25:26:28.12 & {$1.88^{+0.35}_{-0.36}$} & 2.02 & 6.29 & 4.2 & 11.7 \\ 
21:27:48.05 & 25:14:38.07 & {$0.92^{+0.16}_{-0.17}$} & 0.98 & 5.73 & 1.1 & {$<3.5$} \\ 
21:30:51.28 & 24:51:39.28 & {$0.80^{+0.15}_{-0.18}$} & 0.87 & 5.98 & 11.8 & 47.3 \\ 
21:33:26.51 & 25:22:15.72 & {$0.92^{+0.20}_{-0.22}$} & 1.02 & 5.86 & 7.5 & 28.3 \\ 
21:28:13.38 & 24:48:42.27 & {$0.91^{+0.17}_{-0.18}$} & 0.98 & 5.74 & 3.8 & {$<3.5$} \\ 
21:27:08.59 & 25:19:08.56 & {$0.96^{+0.18}_{-0.18}$} & 1.03 & 5.84 & {$<0.33$}& {$<3.5$} \\ 
21:29:54.40 & 24:54:40.10 & {$0.83^{+0.15}_{-0.17}$} & 0.89 & 5.77 & 5.8 & 21.5 \\ 
21:27:22.57 & 24:52:52.14 & {$0.95^{+0.16}_{-0.18}$} & 1.01 & 5.79 & {$<0.33$}& {$<3.5$} \\ 
21:30:30.76 & 25:16:35.42 & {$0.84^{+0.16}_{-0.18}$} & 0.91 & 5.38 & 4.4 & 11.3 \\ 
21:29:06.54 & 25:06:34.41 & {$0.91^{+0.16}_{-0.17}$} & 0.98 & 5.41 & 1.5 & 4.2 \\ 
21:30:53.15 & 25:15:48.59 & {$1.24^{+0.16}_{-0.17}$} & 1.28 & 5.43 & 1.3 & 3.0 \\ 
21:33:15.39 & 25:28:58.52 & {$3.53^{+0.74}_{-0.79}$} & 3.86 & 5.20 & 37.6 & 39.1 \\ 
14:29:10.22 & 35:29:47.05 & {$20.31^{+0.19}_{-0.19}$} & 20.32 & 104.59 & 26.9 & 24.4 \\ 
14:34:34.22 & 35:10:09.53 & {$14.50^{+0.32}_{-0.33}$} & 14.51 & 47.02 & 36.2 & 69.6 \\ 
14:34:00.30 & 35:11:52.30 & {$8.08^{+0.17}_{-0.18}$} & 8.09 & 45.83 & 10.1 & 12.2 \\ 
14:27:58.74 & 34:59:19.92 & {$6.95^{+0.17}_{-0.18}$} & 6.96 & 39.29 & 5.7 & 6.8 \\ 
14:26:32.19 & 35:08:14.88 & {$5.49^{+0.16}_{-0.18}$} & 5.50 & 31.69 & 30.1 & 95.0 \\ 
14:25:41.98 & 34:58:39.37 & {$5.68^{+0.31}_{-0.32}$} & 5.71 & 23.23 & 35.6 & 154.7 \\ 
14:25:40.39 & 34:58:05.78 & {$7.94^{+0.34}_{-0.34}$} & 7.98 & 20.96 & 29.2 & 67.4 \\ 
14:31:34.55 & 35:15:11.30 & {$2.92^{+0.16}_{-0.17}$} & 2.94 & 20.24 & 21.2 & 76.0 \\ 
14:34:13.86 & 34:55:05.49 & {$3.34^{+0.18}_{-0.19}$} & 3.37 & 18.35 & 2.1 & 4.8 \\ 
14:27:59.15 & 34:55:21.07 & {$3.79^{+0.17}_{-0.19}$} & 3.80 & 13.44 & 7.6 & 34.1 \\ 
14:33:15.26 & 34:51:41.46 & {$3.17^{+0.21}_{-0.23}$} & 3.20 & 16.26 & 23.2 & 90.8 \\ 
14:33:20.62 & 34:50:34.85 & {$1.83^{+0.25}_{-0.26}$} & 1.90 & 9.97 & 18.9 & 90.6 \\ 
14:27:02.24 & 34:59:04.28 & {$2.75^{+0.17}_{-0.18}$} & 2.78 & 15.73 & 8.2 & 17.7 \\ 
14:27:41.14 & 34:59:31.52 & {$2.53^{+0.17}_{-0.17}$} & 2.56 & 14.61 & 1.6 & {$<3.5$} \\ 
14:29:09.28 & 35:03:25.16 & {$1.70^{+0.17}_{-0.18}$} & 1.74 & 13.25 & 7.8 & 46.6\tablenotemark{$\dagger$}\\ 
14:29:11.11 & 35:03:20.78 & {$1.08^{+0.17}_{-0.18}$} & 1.14 & 10.44 & 2.7 & 46.6\tablenotemark{$\dagger$}\\ 
14:29:06.01 & 35:11:15.69 & {$2.30^{+0.17}_{-0.19}$} & 2.33 & 13.20 & 6.2 & 13.2 \\ 
14:28:50.46 & 34:54:20.89 & {$2.41^{+0.18}_{-0.19}$} & 2.44 & 12.98 & 5.8 & 9.5 \\ 
14:32:45.57 & 34:55:38.10 & {$2.11^{+0.17}_{-0.17}$} & 2.14 & 12.51 & 0.6 & {$<3.5$} \\ 
14:33:09.89 & 35:15:18.21 & {$1.82^{+0.17}_{-0.17}$} & 1.86 & 11.63 & 9.4 & 29.2 \\ 
14:30:26.60 & 35:19:20.71 & {$3.41^{+0.16}_{-0.17}$} & 3.43 & 11.03 & 5.8 & 20.8 \\ 
14:32:13.54 & 35:09:41.03 & {$1.72^{+0.16}_{-0.17}$} & 1.76 & 10.12 & 6.6 & 16.5 \\ 
14:29:22.90 & 35:12:19.80 & {$1.73^{+0.17}_{-0.18}$} & 1.77 & 10.48 & 2.0 & {$<3.5$} \\ 
14:34:35.36 & 35:26:22.24 & {$3.01^{+0.31}_{-0.32}$} & 3.08 & 10.08 & 14.3 & 30.8 \\ 
14:27:17.93 & 35:01:30.76 & {$1.75^{+0.17}_{-0.18}$} & 1.79 & 9.98 & 3.1 & 4.9 \\ 
14:30:11.82 & 35:00:20.16 & {$1.80^{+0.16}_{-0.17}$} & 1.83 & 9.97 & 5.4 & 12.7 \\ 
14:26:51.48 & 35:19:24.68 & {$1.59^{+0.16}_{-0.17}$} & 1.62 & 9.49 & 5.7 & 4.4 \\ 
14:29:53.81 & 35:17:54.52 & {$1.92^{+0.16}_{-0.17}$} & 1.96 & 9.36 & 0.9 & {$<3.5$} \\ 
14:32:39.93 & 35:11:58.74 & {$1.48^{+0.17}_{-0.17}$} & 1.53 & 8.92 & 0.8 & {$<3.5$} \\ 
14:32:37.98 & 35:30:36.88 & {$1.79^{+0.20}_{-0.21}$} & 1.84 & 8.88 & 7.0 & 19.9 \\ 
14:28:50.71 & 34:53:15.03 & {$1.72^{+0.19}_{-0.21}$} & 1.76 & 8.65 & 2.1 & 2.3 \\ 
14:34:21.94 & 35:25:17.88 & {$1.64^{+0.20}_{-0.22}$} & 1.69 & 8.48 & 4.5 & 9.1 \\ 
14:34:39.66 & 35:08:27.84 & {$3.43^{+0.38}_{-0.40}$} & 3.52 & 8.16 & 2.8 & 5.6 \\ 
14:34:34.17 & 35:09:58.11 & {$2.22^{+0.30}_{-0.32}$} & 2.31 & 8.04 & 36.2 & 69.6 \\ 
14:31:12.34 & 35:35:26.58 & {$4.12^{+0.50}_{-0.52}$} & 4.25 & 7.57 & 3.7 & 6.8 \\ 
14:32:56.09 & 35:33:39.50 & {$2.25^{+0.32}_{-0.34}$} & 2.35 & 7.90 & 17.5 & 66.3 \\ 
14:27:26.98 & 35:02:12.07 & {$1.26^{+0.17}_{-0.19}$} & 1.31 & 7.69 & 0.6 & {$<3.5$} \\ 
14:28:51.10 & 35:30:30.07 & {$1.39^{+0.20}_{-0.22}$} & 1.45 & 7.24 & 7.1 & 23.1 \\ 
14:32:16.89 & 35:02:48.92 & {$1.11^{+0.17}_{-0.18}$} & 1.17 & 7.28 & 0.7 & {$<3.5$} \\ 
14:28:25.47 & 34:55:47.06 & {$1.08^{+0.17}_{-0.19}$} & 1.13 & 7.05 & 4.8 & 11.7 \\ 
14:29:23.66 & 35:28:51.22 & {$1.26^{+0.18}_{-0.19}$} & 1.32 & 6.89 & 2.9 & 7.9 \\ 
14:26:34.41 & 34:59:29.06 & {$0.87^{+0.17}_{-0.18}$} & 0.94 & 5.60 & 3.6 & 8.1 \\ 
14:34:32.42 & 35:22:20.90 & {$1.53^{+0.26}_{-0.28}$} & 1.63 & 6.19 & {$<0.33$}& {$<3.5$} \\ 
14:30:09.94 & 35:19:57.93 & {$0.90^{+0.16}_{-0.18}$} & 0.96 & 6.02 & 0.4 & {$<3.5$} \\ 
14:29:48.68 & 35:17:47.91 & {$1.03^{+0.17}_{-0.17}$} & 1.09 & 6.12 & 5.2 & 15.2 \\ 
14:29:05.66 & 34:49:10.32 & {$1.84^{+0.34}_{-0.38}$} & 1.98 & 5.82 & 11.7 & 33.4 \\ 
14:33:01.39 & 35:01:49.41 & {$0.82^{+0.16}_{-0.19}$} & 0.89 & 5.10 & 0.7 & {$<3.5$} \\ 
14:32:59.91 & 35:28:33.87 & {$0.85^{+0.17}_{-0.19}$} & 0.93 & 5.41 & 5.6 & 17.8 \\ 
14:33:08.12 & 35:31:50.30 & {$1.34^{+0.24}_{-0.26}$} & 1.43 & 5.83 & 1.3 & 2.8 \\ 
14:31:00.42 & 35:36:32.76 & {$5.35^{+0.74}_{-0.76}$} & 5.57 & 5.56 & 12.8 & 38.1 \\ 
14:32:39.56 & 35:01:51.23 & {$0.85^{+0.16}_{-0.19}$} & 0.92 & 5.55 & 6.0 & 15.1 \\ 
14:28:34.05 & 34:57:02.02 & {$0.84^{+0.17}_{-0.19}$} & 0.91 & 5.39 & {$<0.33$}& {$<3.5$} \\ 
14:34:03.67 & 35:24:53.09 & {$0.82^{+0.17}_{-0.18}$} & 0.89 & 5.32 & {$<0.33$}& {$<3.5$} \\ 
14:29:34.77 & 35:27:42.12 & {$0.78^{+0.18}_{-0.19}$} & 0.87 & 5.02 & 0.6 & 2.6 \\ 
02:18:42.20 & 31:49:26.88 & {$147.47^{+0.22}_{-0.22}$} & 147.47 & 662.28 & 112.0 & 56.8 \\ 
02:14:15.63 & 32:03:50.44 & {$26.83^{+0.19}_{-0.20}$} & 26.83 & 131.45 & 30.6 & 11.7 \\ 
02:13:39.17 & 31:59:27.36 & {$2.37^{+0.16}_{-0.18}$} & 2.39 & 23.85 & 11.4 & 62.8\tablenotemark{$\dagger$}\\ 
02:13:38.65 & 31:59:07.72 & {$2.67^{+0.17}_{-0.17}$} & 2.69 & 21.09 & 2.2 & 62.8\tablenotemark{$\dagger$}\\ 
02:13:38.33 & 31:58:48.36 & {$2.12^{+0.17}_{-0.17}$} & 2.15 & 23.64 & 7.6 & {$<3.5$} \\ 
02:16:35.38 & 32:00:22.55 & {$3.26^{+0.15}_{-0.17}$} & 3.28 & 21.08 & 19.7 & 47.5 \\ 
02:14:02.07 & 32:06:13.47 & {$1.90^{+0.17}_{-0.19}$} & 1.94 & 19.77 & 0.9 & {$<3.5$} \\ 
02:14:02.15 & 32:06:06.63 & {$1.91^{+0.18}_{-0.18}$} & 1.95 & 19.30 & 0.6 & {$<3.5$} \\ 
02:16:45.58 & 32:03:58.85 & {$3.20^{+0.16}_{-0.16}$} & 3.21 & 16.59 & 10.4 & 19.5 \\ 
02:18:50.64 & 32:03:28.96 & {$2.68^{+0.16}_{-0.16}$} & 2.70 & 16.43 & 5.3 & 9.8 \\ 
02:19:02.96 & 31:59:47.82 & {$2.11^{+0.16}_{-0.17}$} & 2.14 & 10.50 & 9.8 & 27.5 \\ 
02:13:55.80 & 31:52:59.74 & {$1.86^{+0.17}_{-0.18}$} & 1.89 & 10.36 & 3.7 & 10.0 \\ 
02:15:19.89 & 31:52:42.50 & {$1.66^{+0.18}_{-0.20}$} & 1.71 & 9.53 & 8.9 & 19.1 \\ 
02:16:01.36 & 31:48:22.72 & {$2.56^{+0.29}_{-0.30}$} & 2.63 & 9.23 & 8.7 & 12.7 \\ 
02:17:53.59 & 31:50:14.93 & {$1.69^{+0.20}_{-0.21}$} & 1.75 & 8.84 & 1.5 & {$<3.5$} \\ 
02:12:56.83 & 32:02:22.54 & {$1.34^{+0.16}_{-0.17}$} & 1.39 & 8.41 & 2.7 & 5.7 \\ 
02:14:09.72 & 32:06:39.97 & {$0.86^{+0.19}_{-0.20}$} & 0.95 & 7.00 & 3.9 & 24.5 \\ 
02:13:31.80 & 31:58:08.46 & {$1.10^{+0.16}_{-0.17}$} & 1.16 & 6.84 & 0.4 & {$<3.5$} \\ 
02:12:43.34 & 31:58:56.93 & {$0.96^{+0.16}_{-0.18}$} & 1.02 & 5.76 & 4.3 & 9.6 \\ 
02:15:53.67 & 32:03:03.79 & {$1.01^{+0.16}_{-0.18}$} & 1.07 & 6.21 & 1.7 & 3.0 \\ 
02:17:19.14 & 31:58:11.28 & {$0.80^{+0.16}_{-0.17}$} & 0.88 & 5.42 & -- & {$<3.5$} \\ 
02:18:02.31 & 32:02:59.07 & {$0.79^{+0.16}_{-0.18}$} & 0.87 & 5.11 & -- & {$<3.5$} \\ 
02:22:48.81 & 33:57:19.00 & {$12.35^{+0.13}_{-0.15}$} & 12.35 & 87.84 & -- & 189.3 \\ 
02:22:22.79 & 34:00:15.65 & {$5.63^{+0.51}_{-0.52}$} & 5.74 & 10.98 & -- & 5.5 \\ 
02:22:46.21 & 34:04:33.89 & {$1.39^{+0.31}_{-0.34}$} & 1.54 & 5.18 & -- & 15.1 \\ 
02:30:41.98 & 33:56:21.46 & {$4.02^{+0.20}_{-0.20}$} & 4.04 & 20.28 & -- & 35.3 \\ 
02:11:40.54 & 33:04:17.31 & {$2.59^{+0.30}_{-0.32}$} & 2.67 & 8.56 & -- & 21.4 \\ 
02:20:45.66 & 32:57:05.64 & {$3.13^{+0.24}_{-0.24}$} & 3.18 & 8.50 & -- & 23.5 \\ 
02:24:36.11 & 32:51:45.89 & {$4.43^{+0.30}_{-0.31}$} & 4.48 & 14.09 & -- & 8.5 \\ 
02:16:03.95 & 33:03:45.08 & {$1.03^{+0.21}_{-0.23}$} & 1.12 & 5.27 & -- & {$<3.5$} \\ 
02:20:48.03 & 32:41:06.53 & {$204.05^{+0.47}_{-0.49}$} & 204.05 & 417.06 & -- & 921.6 \\ 
02:20:17.28 & 32:38:53.09 & {$4.28^{+0.26}_{-0.27}$} & 4.32 & 16.28 & 10.1 & 10.2 \\ 
02:25:00.23 & 32:31:36.95 & {$7.41^{+0.36}_{-0.36}$} & 7.45 & 20.48 & -- & 16.8 \\ 
02:24:40.90 & 32:29:39.65 & {$1.21^{+0.27}_{-0.29}$} & 1.33 & 5.16 & -- & {$<3.5$} \\ 
02:24:27.44 & 32:34:27.52 & {$0.83^{+0.13}_{-0.14}$} & 0.87 & 4.72 & -- & 2.3 \\ 
02:11:48.05 & 32:09:32.41 & {$1.20^{+0.26}_{-0.28}$} & 1.32 & 5.05 & -- & {$<3.5$} \\ 
02:20:24.90 & 32:10:45.30 & {$1.45^{+0.11}_{-0.13}$} & 1.46 & 11.18 & 7.3 & 17.5 \\ 
02:20:11.45 & 32:10:42.50 & {$1.46^{+0.19}_{-0.21}$} & 1.52 & 6.36 & 5.9 & 19.7 \\ 
02:24:40.70 & 32:02:05.45 & {$6.20^{+0.33}_{-0.33}$} & 6.24 & 19.66 & -- & 6.5 \\ 
02:24:40.56 & 32:08:43.00 & {$1.25^{+0.13}_{-0.13}$} & 1.28 & 9.94 & -- & 7.1 \\ 
02:24:59.44 & 32:14:04.83 & {$2.24^{+0.28}_{-0.29}$} & 2.32 & 6.30 & -- & 25.3 \\ 
02:16:27.02 & 32:08:15.51 & {$0.76^{+0.12}_{-0.15}$} & 0.81 & 6.27 & 0.7 & {$<3.5$} \\ 
02:16:28.14 & 32:10:30.20 & {$0.62^{+0.12}_{-0.14}$} & 0.67 & 5.48 & 1.6 & 3.4 \\ 
14:18:40.52 & 35:30:02.51 & {$6.82^{+0.14}_{-0.15}$} & 6.82 & 47.72 & -- & 91.5\tablenotemark{$\dagger$}\\ 
14:18:40.59 & 35:30:23.21 & {$0.85^{+0.16}_{-0.17}$} & 0.91 & 5.88 & -- & 91.5\tablenotemark{$\dagger$}\\ 
14:18:40.27 & 35:29:41.86 & {$2.53^{+0.21}_{-0.21}$} & 2.57 & 10.89 & -- & {$<3.5$} \\ 
14:18:40.86 & 35:33:24.71 & {$1.38^{+0.18}_{-0.18}$} & 1.44 & 8.22 & -- & 27.4 \\ 
14:22:54.68 & 35:32:08.97 & {$0.72^{+0.14}_{-0.16}$} & 0.78 & 5.28 & -- & 3.4 \\ 
14:22:42.74 & 35:31:48.44 & {$0.68^{+0.15}_{-0.17}$} & 0.75 & 5.05 & -- & 4.5 \\ 
21:24:22.68 & 24:55:12.95 & {$8.72^{+0.30}_{-0.31}$} & 8.74 & 29.65 & -- & 4.5\\ 
21:24:32.61 & 25:03:17.39 & {$1.75^{+0.32}_{-0.34}$} & 1.87 & 5.31 & -- & 5.8 \\ 
14:22:38.77 & 34:58:18.65 & {$1.37^{+0.24}_{-0.26}$} & 1.47 & 6.00 & -- & {$<3.5$} \\ 
21:25:03.78 & 25:25:12.85 & {$4.58^{+0.54}_{-0.56}$} & 4.72 & 8.56 & -- & {$<3.5$} \\ 
21:24:34.23 & 25:27:58.04 & {$0.98^{+0.19}_{-0.20}$} & 1.06 & 4.41 & -- & 7.4 \\ 
02:11:43.10 & 33:31:39.73 & {$4.27^{+0.21}_{-0.23}$} & 4.29 & 19.54 & -- & 12.6 \\ 
02:11:31.56 & 33:23:02.49 & {$4.22^{+0.28}_{-0.30}$} & 4.26 & 14.80 & -- & {$<3.5$} \\ 
02:19:30.26 & 33:28:51.69 & {$57.66^{+0.25}_{-0.26}$} & 57.67 & 224.25 & -- & 100.7 \\ 
02:19:43.75 & 33:30:28.66 & {$1.17^{+0.20}_{-0.22}$} & 1.25 & 6.10 & -- & {$<3.5$} \\ 
02:24:23.34 & 33:32:30.01 & {$3.23^{+0.29}_{-0.30}$} & 3.29 & 11.45 & -- & 18.3 \\ 
02:23:33.81 & 33:24:35.41 & {$3.91^{+0.60}_{-0.62}$} & 4.11 & 6.77 & -- & 54.5 \\ 
21:24:50.28 & 26:01:50.36 & {$7.85^{+0.19}_{-0.20}$} & 7.86 & 33.37 & -- & 8.6\tablenotemark{$\dagger$}\\ 
21:24:50.34 & 26:01:59.06 & {$1.85^{+0.21}_{-0.22}$} & 1.91 & 8.50 & -- & 8.6\tablenotemark{$\dagger$}\\ 
21:33:02.14 & 26:04:17.99 & {$1.32^{+0.29}_{-0.32}$} & 1.46 & 5.24 & -- & {$<3.5$} \\ 
21:36:42.22 & 25:57:41.47 & {$15.21^{+0.50}_{-0.52}$} & 15.24 & 28.20 & -- & 10.4 \\ 
21:36:55.58 & 25:56:43.63 & {$4.38^{+0.30}_{-0.31}$} & 4.42 & 14.43 & -- & 6.1 \\ 
\enddata
\label{tab:srcs}
\end{longtable}

\begin{longtable}{ccc}
\tabletypesize{\scriptsize}
\tablecolumns{2}
\tablewidth{0pt}
\tablecaption{SZA Area Coverage}
\tablehead{
\colhead{Noise Value} & \colhead{$\sum$ Area} \\
 \colhead{(mJy)} & \colhead{($\deg^2$)}
 }
 \startdata
0.11 & 0.000 \\ 
0.12 & 0.014 \\ 
0.13 & 0.037 \\ 
0.14 & 0.080 \\ 
0.15 & 0.219 \\ 
0.16 & 0.638 \\ 
0.17 & 2.578 \\ 
0.18 & 3.629 \\ 
0.19 & 4.012 \\ 
0.20 & 4.281 \\ 
0.21 & 4.504 \\ 
0.22 & 4.697 \\ 
0.23 & 4.868 \\ 
0.24 & 5.023 \\ 
0.26 & 5.323 \\ 
0.28 & 5.554 \\ 
0.30 & 5.758 \\ 
0.32 & 5.940 \\ 
0.34 & 6.106 \\ 
0.36 & 6.257 \\ 
0.38 & 6.396 \\ 
0.40 & 6.525 \\ 
0.45 & 6.807 \\ 
0.50 & 7.036 \\ 
0.55 & 7.226 \\ 
0.60 & 7.385 \\ 
0.65 & 7.514 \\ 
0.70 & 7.613 \\ 
0.75 & 7.684 \\ 
\enddata
\label{tab:area}
\end{longtable}

\label{app:sources}

\end{document}